\documentclass[aps,prd,10pt,superscriptaddress,a4paper,twocolumn]{revtex4-2}
\usepackage{blindtext}
\usepackage{graphicx}
\usepackage{amssymb}
\usepackage{color}
\usepackage{amsmath}
\usepackage{empheq}
\usepackage{lipsum}
\usepackage{orcidlink}
\usepackage{physics}
\usepackage{mathrsfs}
\usepackage{MnSymbol}
\usepackage{comment}

\newcommand{\be}{\begin{equation}}
\newcommand{\ee}{\end{equation}}

\newcommand{\PR}[1]{\ensuremath{\left[#1\right]}}
\newcommand{\PC}[1]{\ensuremath{\left(#1\right)}}

\begin{document}
\title{Scalar fields, impurities and supersymmetry}

\author{D. Bazeia
\footnote{Email: dbazeia@gmail.com}
\orcidlink{0000-0003-1335-3705}}
\affiliation{{Departamento de Física, Universidade Federal da Paraíba, 58051-970 João Pessoa, Paraíba, Brazil}}

\author{A.C. Lehum
\footnote{Email: lehum@ufpa.br}
\orcidlink{0000-0001-5075-6541}}
\affiliation{{Faculdade de Física, Universidade Federal do Pará, 66075-110, Belém, Pará, Brazil}}

\author{G.S. Santiago
\footnote{Email: gss.santiago99@gmail.com}
\orcidlink{0009-0006-0529-4708}}
\affiliation{{Departamento de Física, Universidade Federal da Paraíba, 58051-970 João Pessoa, Paraíba, Brazil}}

\begin{abstract}
We develop a rigid $\mathcal{N}=(1,1)$ superspace formulation for multifield scalar models coupled to localized impurities through spurion superfields in two-dimensional spacetime. The spurionic completion gives a manifestly supersymmetric action and provides a systematic framework for describing interacting scalar fields in the presence of impurity backgrounds. In this setting, supersymmetry acts as an organizing principle for a controlled bosonic half-BPS sector, with the preserved projector selecting the impurity-compatible first-order equations. We derive the corresponding coupled BPS equations, energy density, boundary conditions, and Bogomol'nyi bound, showing that localized impurities deform the BPS profiles and redistribute the local energy density while leaving the total BPS energy fixed by the topological boundary term. We illustrate the formalism through representative one-, two-, and three-field models, analyzing the resulting BPS configurations and their local energy density profiles.
\end{abstract}

\maketitle

\section{Introduction}
The search for localized structures described by real scalar fields in $1+1$ spacetime dimensions started long ago, engendering distinct motivations, in particular, in high energy and condensed matter physics. Some of the first investigations were described in Refs. \cite{A,B,C,D,E,F}, and further considered in the books \cite{B1,B2,B3,B4,B5}, where examples of current interest are explicitly described and analyzed. 
In the presence of impurities, the study of localized structures also started long ago; see, e.g., Refs. \cite{I1,I2,I3} and references therein.
Here, one of the basic motivations for dealing with more realistic models is the use of impurities to describe doping, point defects, or imperfections in the system. Well-known possibilities concern Anderson localization \cite{AL}, where the presence of impurities may significantly alter the physical behavior of the system, and also the Kondo effect, where the magnetic resistance of a metal may be altered by the presence of magnetic impurities \cite{KE}.

Several of the first studies dealing with impurities have considered the case where inhomogeneities appear as local or point-like modifications of the system under investigation. More recently, however, results have appeared where the inhomogeneity is itself a localized but not point-like structure. A particularly interesting study of localized structures in the presence of localized impurities appeared earlier in Ref. \cite{Adam1}. There, the phenomenon of the spectral wall was investigated and shown to occur in a model with a BPS preserving impurity, once the static force between the antikink and the impurity vanishes. The result motivated the authors of Ref. \cite{Adam2} to further consider models of scalar fields in the presence of impurities within the supersymmetric environment. On the other hand, previous investigations of scalar fields in the presence of impurities have appeared in Refs. \cite{Adam3,Adam4,Liao1,Liao2,Liao3}, which motivated another possibility involving supersymmetry in the presence of a real spurion superfield \cite{Lehum}, that seems to provide a very adequate supersymmetric description to motivate extensions retaining supersymmetric control over inhomogeneous deformations. In this context, in this work we follow \cite{Lehum} and focus on an extension to the case of a multifield model containing several real scalar fields, together with the accompanying fermionic degrees of freedom. As we will show below, the multifield construction allows the impurity background to affect the different scalars in distinct ways, making it possible to generate impurity-induced effects that finds no counterpart in the single-field framework. Furthermore, the extra scalar fields may be important to broaden the scope of the system; this occurs, for instance, in the case of magnetic materials, in the study of magnetic Néel and Bloch walls, and skymions \cite{AAAB,AAAa,AAAR}. The added scalars can also introduce other effects, as the construction of intersecting domain walls \cite{BBB,Saffin} and multifield spectral walls \cite{Adam4}. The presence of fermions opens up additional possibilities, particularly the construction of kink–impurity models within a manifestly supersymmetric framework. This provides a straightforward procedure for establishing a first-order formalism for coupled scalar fields.
Another motivation is that the addition of other fields may be an important step towards including vector fields and working in higher spatial dimensions, a subject to be studied elsewhere. 

To implement the present investigation, in the next Section we introduce the methodology. There, the general model is formulated in rigid ${\cal N}=(1,1)$ superspace for $N$ real matter superfields coupled to $N$ real spurion superfields. After deriving the off-shell component Lagrangian, we identify the half-BPS sector selected by static impurity backgrounds and obtain the corresponding first-order equations, energy density, boundary conditions, and Bogomol'nyi bound. Moreover, in Sec. \ref{sec3} we illustrate the main results by investigating several models described by one, two, and three real scalar fields in the presence of distinct impurities. We show that the generalized spurionic construction provides a controlled framework for obtaining analytical and numerical half-BPS solutions in impurity-deformed scalar models, including configurations with impurity-induced internal structures and regions of negative energy density. As a particularly interesting result, in suitable cases, the impurity-deformed first-order equations can also be mapped into the corresponding impurity-free BPS system through a reparametrization of the spatial coordinate. We end the work in Sec. \ref{sec4}, highlighting the main results and suggesting new lines for further research.

\section{Methodology} 

Let us focus on the search for localized structures in models described by scalar fields in the presence of impurities within a supersymmetric framework. Here we consider an extension of the construction recently proposed in Ref. \cite{Lehum} to a system composed of $N$ matter superfields $\Phi_{i}$ and $N$ spurion superfields $\Sigma_{i}$. This extension is of current interest and may also serve as a useful guide toward further developments, including the incorporation of Abelian and non-Abelian vector fields, as well as extensions to one and two additional spatial dimensions.

In two-dimensional spacetime, we consider a rigid $\mathcal{N} = (1,1)$ superspace over a spacetime with metric signature $(-,+)$. The full superspace action is taken to be
\begin{equation}
\begin{aligned}
    S = \int d^2xd^2\theta\bigg(&-\frac{1}{4}D^{\alpha}\Phi_{i}D_{\alpha}\Phi_{i} - W(\Phi_{i})\\
    &- \Sigma_{i}\mathcal{F}_{i}(\Phi_{j},\Sigma_{k}) - \frac{1}{2}\Sigma_{i}\Sigma_{i}\bigg).
\end{aligned}
\end{equation}
Throughout this work, we use the superspace conventions of Ref.~\cite{GGRS}. Repeated indices are understood to be summed over. Latin indices $i,j,k,\ldots=1,\ldots,N$ label the matter and spurion superfields, whereas Greek indices $\alpha,\beta,\ldots=\pm$ denote spinor components. We assume that $W(\Phi_{i})$ and $\mathcal{F}_{i}(\Phi_{j},\Sigma_{k})$ are arbitrary real scalar superfunctions. The $N$ real matter superfields $\Phi_{i}$ and the $N$ real spurion superfields $\Sigma_{i}$ are expanded in powers of $\theta$ as
\begin{subequations}
\begin{align}
    \Phi_{i} &= \phi_{i} + \theta^{\beta}\psi_{i\beta} - \theta^{2}F_{i},\\
    \Sigma_{i} &= \sigma_{i} + \theta^{\beta}\lambda_{i\beta} - \theta^{2}G_{i}.
\end{align}
\end{subequations}
The components $\psi_{i}^{\beta}$ and $\lambda_{i}^{\beta}$ are real Majorana-Weyl fermions, whereas $F_{i}$ and $G_{i}$ are real auxiliary scalar fields. We emphasize that the spurion superfields $\Sigma_{i}$ are treated as fixed background superfields and, therefore, are not varied in the extremization of the action. To evaluate the kinetic term, we expand
\begin{subequations}
\begin{align}
    D_{\alpha}\Phi_{i} &= \psi_{i\alpha} + \theta^{\mu}\PC{i\partial_{\mu\alpha}\phi_{i} - C_{\mu\alpha}F_{i}} - \theta^{2}i\partial_{\mu\alpha}\psi_{i}^{\mu},\\
    D^{\alpha}\Phi_{i} &= \psi_{i}^{\alpha} + \theta^{\mu}i\partial_{\mu}^{\alpha}\phi_{i} - \theta^{\alpha}F_{i} - \theta^{2}i\partial_{\mu}^{\alpha}\psi_{i}^{\mu},
\end{align}
\end{subequations}
where $C_{\alpha\beta}$ is a second-rank antisymmetric tensor used to raise and lower spinor indices. By multiplying these expressions and extracting only the $\theta^{2}$ component, we obtain
\begin{equation}
    \PR{D^{\alpha}\Phi_{i}D_{\alpha}\Phi_{i}}_{\theta^{2}} = 2\PC{\phi_{i}\Box\phi_{i} + \psi_{i\alpha}i\partial_{\mu}^{\alpha}\psi_{i}^{\mu}+F_{i}^{2}}.
\end{equation}

The kinetic contribution to the action therefore reads
\begin{equation}
    S_{\textrm{kin}} = \frac{1}{2}\int d^{2}x\PC{\phi_{i}\Box\phi_{i} + \psi_{i\alpha}i\partial_{\mu}^{\alpha}\psi_{i}^{\mu}+F_{i}^{2}}.
\end{equation}
Analogously, expanding the real scalar superfunction $W(\Phi_{i})$ and extracting its $\theta^{2}$ component gives
\begin{equation}
    \PR{W(\Phi_{i})}_{\theta^{2}} = W_{\phi_{l}}F_{l} + \frac{1}{2}W_{\phi_{l}\phi_{p}}\psi_{p}^{\alpha}\psi_{l\alpha},
\end{equation}
where $W_{\phi_{l}} = \partial W/\partial\phi_{l}$, $W_{\phi_{l}\phi_{p}} = \partial^{2}W/\PC{\partial\phi_{l}\partial\phi_{p}}$, with analogous notation for higher derivatives. Similarly, the $\theta^{2}$ component of the coupling term is
\begin{equation}\label{eq_comp_spurion}
\begin{aligned}
    &\PR{\Sigma_{i}\mathcal{F}_{i}(\Phi_{j},\Sigma_{k})}_{\theta^{2}} = G_{i}\mathcal{F}_{i} \!+\! \mathcal{F}_{i\phi_{l}}\lambda_{i}^{\alpha}\psi_{l\alpha} \!+\! \mathcal{F}_{i\sigma_{l}}\lambda_{i}^{\alpha}\lambda_{l\alpha}\\
    &\!+\! \sigma_{i}\mathcal{F}_{i\phi_{l}\sigma_{p}}\lambda_{p}^{\alpha}\psi_{l\alpha} \!+\! \frac{\sigma_{i}}{2}\!\PC{\!\mathcal{F}_{i\phi_{l}\phi_{p}}\psi_{p}^{\alpha}\psi_{l\alpha} \!+\! \mathcal{F}_{i\sigma_{l}\sigma_{p}}\lambda_{p}^{\alpha}\lambda_{l\alpha}}\\
    &+\sigma_{i}\mathcal{F}_{i\phi_{l}}F_{l} + \sigma_{i}\mathcal{F}_{i\sigma_{l}}G_{l},
\end{aligned}
\end{equation}
while the quadratic spurion term gives
\begin{equation}
    \PR{\frac{1}{2}\Sigma_{i}\Sigma_{i}}_{\theta^{2}} = \sigma_{i}G_{i} + \frac{1}{2}\lambda_{i}^{\alpha}\lambda_{i\alpha}.
\end{equation}

Collecting the above contributions, the full off-shell component Lagrangian is given by
\begin{equation}
\begin{aligned}
\label{Loff}
    &\mathcal{L}_{\textrm{off}} = \frac{1}{2}\phi_{i}\Box\phi_{i} + \frac{i}{2}\PC{\psi_{i}^{+}\partial_{++}\psi_{i}^{+}+\psi_{i}^{-}\partial_{--}\psi_{i}^{-}}\\
    &\!+\!\frac{1}{2}\!F_{i}^2 \!-\! F_{i}\PC{W_{\phi_{i}} \!+\! \sigma_{l}\mathcal{F}_{l\phi_{i}}\!}\\
    &-\!\frac{i}{2}\!\PC{W_{\phi_{l}\phi_{p}}\!+\!\sigma_{i}\mathcal{F}_{i\phi_{l}\phi_{p}}\!}\PC{\psi_{p}^{+}\!\psi_{l}^{-} - \psi_{p}^{-}\!\psi_{l}^{+}}\\
    &\!-\! i\PC{\mathcal{F}_{p\phi_{l}} + \sigma_{i}\mathcal{F}_{i\phi_{l}\sigma_{p}}}\!\PC{\lambda_{p}^{+}\psi_{l}^{-}\! - \!\lambda_{p}^{-}\psi_{l}^{+}}\\
    &-i\PC{\mathcal{F}_{p\sigma_{l}} + \frac{1}{2}\sigma_{i}\mathcal{F}_{i\sigma_{l}\sigma_{p}}}\PC{\lambda_{p}^{+}\lambda_{l}^{-} - \lambda_{p}^{-}\lambda_{l}^{+}}\\
    &-G_{i}\PC{\mathcal{F}_{i}+\sigma_{i} + \sigma_{l}\mathcal{F}_{l\sigma_{i}}}-i\lambda_{i}^{+}\lambda_{i}^{-}.
\end{aligned}
\end{equation}
Varying this Lagrangian with respect to the auxiliary fields $F_{i}$ yields the $N$ algebraic equations
\begin{equation}
\label{Frel}
    F_{i} = W_{\phi_{i}} + \sigma_{l}\mathcal{F}_{l\phi_{i}}.
\end{equation}
Under infinitesimal supersymmetry transformations, the components of the matter multiplet transform as follows
\begin{subequations}
\begin{align}
    \delta\phi_{i} &= -i\PC{\varepsilon^{+}\psi_{i}^{-} - \varepsilon^{-}\psi_{i}^{+}},\\
    \delta\psi_{i-} & = -i\PC{\varepsilon^{+}F_{i} + \varepsilon^{-}\partial_{--}\phi_{i}},\\
    \delta\psi_{i+} &= i\PC{\varepsilon^{-}F_{i} - \varepsilon^{+}\partial_{++}\phi_{i}},\\
    \delta F_{i} &= i\PC{\varepsilon^{+}\partial_{++}\psi_{i}^{+}+ \varepsilon^{-}\partial_{--}\psi_{i}^{-}}.
\end{align}
\end{subequations}
Similarly, for the spurion multiplet one obtains
\begin{subequations}
\begin{align}
    \delta\sigma_{i} &= -i\PC{\varepsilon^{+}\lambda_{i}^{-} - \varepsilon^{-}\lambda_{i}^{+}},\\
    \delta\lambda_{i-} &= -i\PC{\varepsilon^{+}G_{i} + \varepsilon^{-}\partial_{--}\sigma_{i}},\\
    \delta\lambda_{i+} &= i\PC{\varepsilon^{-}G_{i} - \varepsilon^{+}\partial_{++}\sigma_{i}},\\
    \delta G_{i} &= i\PC{\varepsilon^{+}\partial_{++}\lambda_{i}^{+} + \varepsilon^{-}\partial_{--}\lambda_{i}^{-}}.
\end{align}
\end{subequations}

We now consider a static bosonic spurion background
\begin{equation}
    \lambda_{i\pm} = 0, \quad \sigma_{i} = \sigma_{i}(x), \quad G_{i} = G_{i}(x).
\end{equation}
For such a static background, the derivatives of the impurity profiles reduce to
\begin{equation}
    \partial_{++}\sigma_{i} = \sigma_{i}^{\prime}(x), \quad \partial_{--}\sigma_{i} = -\sigma_{i}^{\prime}(x),
\end{equation}
where $\sigma_{i}^{\prime}(x) = d\sigma_{i}/dx$. Requiring the background to be invariant under a residual supersymmetry amounts to imposing $\delta\lambda_{i\pm} = 0$ for at least one nonzero pair of constant supersymmetry parameters $\PC{\varepsilon^{+},\varepsilon^{-}}$. This condition implies that
\begin{subequations}
\begin{align}
    \varepsilon^{+}\sigma_{i}^{\prime} - \varepsilon^{-}G_{i} &=0,\label{dlp}\\
    \varepsilon^{+}G_{i} - \varepsilon^{-}\sigma_{i}^{\prime} &= 0.
\end{align}
\end{subequations}
For each nontrivial spurion background, the compatibility of these equations requires
\begin{equation}
\label{Grel}
    G_{i} = \eta\sigma_{i}^{\prime}, \quad \eta = \pm 1.
\end{equation}
Substituting this equation into Eq. \eqref{dlp} gives corresponding projection condition on the supersymmetry parameters,
\begin{equation}
\label{projection}
    \varepsilon^{-} = \eta\varepsilon^{+}.
\end{equation}
Thus, the preserved supercharge is given by
\begin{equation}
    \mathcal{Q}_{\eta} = Q_{+} + \eta Q_{-}, \quad \delta_{\eta} = \varepsilon^{+}\mathcal{Q}_{\eta}.
\end{equation}
Therefore, the $N$ spatially varying impurities can preserve \textit{half} of the $\mathcal{N} = (1,1)$ supersymmetry, provided that their auxiliary partners satisfy Eq. \eqref{Grel}. All nontrivial spurion backgrounds must, however, be compatible with the same choice of $\eta$. Otherwise, different impurity profiles would impose inequivalent projections on the supersymmetry parameters, and no common half-BPS subalgebra would be preserved.

We next consider a static bosonic matter configuration
\begin{equation}
    \psi_{i\pm} = 0, \quad \phi_{i} = \phi_{i}(x), \quad F_{i} = F_{i}(x).
\end{equation}
In order for this configuration to preserve the same residual supersymmetry selected by the spurion background, we must also impose $\delta\psi_{i\pm} = 0$. From $\delta\psi_{i+} = 0$, we get
\begin{equation}
    \varepsilon^{+}\phi_{i}^{\prime} - \varepsilon^{-}F_{i} = 0.
\end{equation}
Using the projection condition in Eq. \eqref{projection}, together with the auxiliary-field equation $F_{i}=W_{\phi_{i}}+\sigma_{l}\mathcal{F}_{l\phi_{i}}$, this condition becomes
\begin{equation}
    \varepsilon^{+}\PC{\phi_{i}^{\prime} - \eta\PC{W_{\phi_{i}}+\sigma_{l}\mathcal{F}_{l\phi_{i}}}} = 0.
\end{equation}
Since $\varepsilon^{+}\neq 0$, the half-BPS equations are
\begin{equation}
\label{eqBPS}
    \phi_{i}^{\prime} = \eta\PC{W_{\phi_{i}}+\sigma_{l}\mathcal{F}_{l\phi_{i}}}.
\end{equation}
Thus, the sign appearing in the matter first-order equations is not independent: it is fixed by the projector already imposed by the spurion background. Consequently, all $N$ first-order equations must involve the same value of $\eta$. Otherwise, the matter sector would not preserve the same projected supersymmetry as the spurion background.

The system studied in Ref. \cite{Liao3} can be recovered by restricting ourselves to the bosonic sector $(\psi^{\pm} = \lambda^{\pm} =0)$ and selecting $\mathcal{F}_{i}\PC{\Phi_{j},\Sigma_{k}} = \Phi_{i}$, so that, in components, $\mathcal{F}_{i} = \phi_{i}-\theta^2 F_i$. This choice yields 
\begin{equation} \mathcal{F}_{l\phi_{i}} = \frac{\partial\phi_{l}}{\partial\phi_{i}} = \delta^{l}_{i}, \quad \mathcal{F}_{l\sigma_{i}} = \frac{\partial\phi_{l}}{\partial\sigma_{i}} = 0. 
\end{equation} 
It follows that the off-shell Lagrangian density defined in Eq. \eqref{Loff} reduces to 
\begin{equation} 
\begin{aligned} \mathcal{L}_{\textrm{off}} &= \frac{1}{2}\,\phi_{i}\Box\phi_{i} \!+\!\frac{1}{2}\!F_{i}^2 \!-\! F_{i}\PC{W_{\phi_{i}} \!+\! \sigma_{i}\!}\\ &-G_{i}\PC{\phi_{i}+\sigma_{i}}. 
\end{aligned} 
\end{equation} 
Using the relations in Eqs. \eqref{Frel} and \eqref{Grel}, we obtain 
\begin{equation} \mathcal{L}_{\textrm{off}} = \frac{1}{2}\phi_{i}\Box\phi_{i} \!-\!\frac{1}{2} \PC{W_{\phi_{i}} \!+\! \sigma_{i}\!}^2 -\eta\sigma_{i}^{\prime}\PC{\phi_{i}+\sigma_{i}}. 
\end{equation} 
We now use the identities 
\begin{subequations} \label{tdt} 
\begin{align} \eta\sigma_{i}^{\prime}\phi_{i} &= \frac{d}{dx}\PC{\eta\sigma_{i}\phi_{i}} - \eta\sigma_{i}\phi_{i}^{\prime},\\ \sigma_{i}^{\prime}\sigma_{i} &= \frac{1}{2}\frac{d}{dx}\PC{\sigma_{i}^{2}}, 
\end{align} 
\end{subequations} 
and discard the total derivative terms. The Lagrangian density then becomes 
\begin{equation} 
\begin{aligned} \mathcal{L}_{\textrm{off}} &= \frac{1}{2}\phi_{i}\Box\phi_{i} - \frac{1}{2}W_{\phi_{i}}^{2} - W_{\phi_{i}}\sigma_{i} \\ & - \frac{1}{2}\sigma_{i}^{2} + \eta\sigma_{i}\phi_{i}^{\prime}. 
\end{aligned} 
\end{equation} 
This is precisely the bosonic scalar-impurity structure studied in Ref. \cite{Liao3}, now derived from the present supersymmetric spurion construction in the multifield setting.

Let us now focus on the energy density and boundary conditions. In the bosonic sector, the off-shell Lagrangian density defined in Eq. \eqref{Loff} reduces to
\begin{equation} \begin{aligned} \mathcal{L}_{\textrm{off}} &= \frac{1}{2}\,\phi_{i}\Box\phi_{i} \!+\!\frac{1}{2}\!F_{i}^2 \!-\! F_{i}\PC{W_{\phi_{i}} \!+\! \sigma_{l}\mathcal{F}_{l\phi_{i}}\!}\\ &-G_{i}\PC{\mathcal{F}_{i} + \sigma_{i} + \sigma_{l}\mathcal{F}_{l\sigma_{i}}}. \end{aligned} \end{equation} For static configurations, the energy density is $\rho(x) = -\mathcal{L}_{\textrm{off}}\vert_{\textrm{static}}$. This gives
\begin{equation} \begin{aligned} \rho &= \frac{1}{2}\phi_{i}^{\prime}\phi_{i}^{\prime} \!-\!\frac{1}{2}\!F_{i}^2 \!+\! F_{i}\PC{W_{\phi_{i}} \!+\! \sigma_{l}\mathcal{F}_{l\phi_{i}}\!}\\ &+G_{i}\PC{\mathcal{F}_{i} + \sigma_{i} + \sigma_{l}\mathcal{F}_{l\sigma_{i}}}. \end{aligned}
\end{equation}
Substituting the auxiliary-field relations in Eqs. \eqref{Frel} and \eqref{Grel} into the expression above, we find
\begin{equation} \label{rho} \begin{aligned} \rho = \frac{1}{2}\phi_{i}^{\prime}\phi_{i}^{\prime} &+ \frac{1}{2}\PC{W_{\phi_{i}} \!+\! \sigma_{l}\mathcal{F}_{l\phi_{i}}\!}^2 \\ &+ \eta\sigma^{\prime}_{i}\PC{\mathcal{F}_{i} + \sigma_{i} + \sigma_{l}\mathcal{F}_{l\sigma_{i}}}. \end{aligned}
\end{equation}
It should be emphasized that, in contrast with the impurity-free case, the last term $\eta\sigma^{\prime}_{i}\PC{\mathcal{F}_{i} + \sigma_{i} + \sigma_{l}\mathcal{F}_{l\sigma_{i}}}$ is not positive definite. Therefore, the energy density can develop regions with negative values, as is typical of kink-impurity models.

In order to obtain finite-energy localized configurations, the non-total-derivative contributions to the energy density must vanish at spatial infinity. For localized impurities, we impose 
\begin{equation} \lim_{x\to\pm\infty}\sigma_{i}(x) = 0, 
\qquad 
\lim_{x\to\pm\infty}\sigma_{i}^{\prime}(x) = 0. \end{equation} 
This condition also justifies discarding the total derivative terms in Eq. \eqref{tdt}. Assuming that the functions $\mathcal{F}_{i}$ and their relevant derivatives remain finite asymptotically, the terms $\sigma_{l}\mathcal{F}_{l\phi_{i}}$ and $\eta\sigma^{\prime}_{i}\PC{\mathcal{F}_{i} + \sigma_{i} + \sigma_{l}\mathcal{F}_{l\sigma_{i}}}$ vanish at spatial infinity. Therefore, the localized energy boundary conditions reduce to 
\begin{equation} \lim_{x\to\pm\infty}W_{\phi_{i}}(x) = 0, \qquad \lim_{x\to\pm\infty}\phi_{i}^{\prime}(x) = 0. \end{equation} 
Thus, the scalar fields $\phi_{i}(x)$ must approach critical points of the superpotential with vanishing derivatives. In general, if the functions $\sigma_{i}$ are non-localized, the asymptotic field values of finite-energy configurations need not be critical points of the superpotential, and this may modify the usual topological prescription of the solutions.

The energy density can also be rewritten in Bogomol'nyi form as 
\begin{equation} 
\begin{aligned} \rho &= \frac{1}{2}\PC{\phi_{i}^{\prime} - \eta\PC{W_{\phi_{i}} \!+\! \sigma_{l}\mathcal{F}_{l\phi_{i}}\!}}^2 \\ &+ \eta\frac{d}{dx}\PC{W + \sigma_{l}\mathcal{F}_{l} + \frac{1}{2}\sigma_{l}\sigma_{l}}. 
\end{aligned} 
\end{equation} 
The bound is saturated precisely when the square term vanishes, namely when the half-BPS equations \eqref{eqBPS} are satisfied. The corresponding Bogomol'nyi energy $E_{B}$ is therefore
\begin{equation}
E_{B} = \int_{-\infty}^{\infty}dx\PC{\eta\frac{d}{dx}\PC{W + \sigma_{l}\mathcal{F}_{l} + \frac{1}{2}\sigma_{l}\sigma_{l}}}, 
\end{equation} 
which gives 
\begin{equation} E_{B} = \eta\PR{W + \sigma_{l}\mathcal{F}_{l} + \frac{1}{2}\sigma_{l}\sigma_{l}}_{x = -\infty}^{x= \infty}. 
\end{equation} 
Since we are considering localized impurities, the terms $\sigma_{l}\mathcal{F}_{l}$ and $\sigma_{l}\sigma_{l}$ vanish asymptotically. Thus, the energy lower bound reduces to
\begin{equation}
\label{EB}
E_{B} = \eta\Big(W(\phi_{i}(\infty)) - W(\phi_{i}(-\infty))\Big). 
\end{equation}

As in the standard BPS argument, the bound follows from the positivity of suitable combinations of supercharges in the presence of topological charges \cite{Witten:1978mh}. In the present case, the sign $\eta$ is fixed by the preserved supersymmetry projector and is therefore not an independent choice within a given BPS branch. Defining $\Delta W = W(\phi_i(\infty))-W(\phi_i(-\infty))$, the compatible BPS flow satisfies $\eta\,\Delta W\geq0$, so that the saturated energy can be written as $E_{\rm B}=|\Delta W|$. The opposite sign of $\eta$ corresponds to the reversed topological sector. It is also of interest to notice that the localized impurity profiles and associated coupling functions do not alter the energy lower bound, although they influence the local behavior of the energy density.
 

\section{Illustration}\label{sec3}

\subsection{One-Field Models}
As recently developed in Ref. \cite{Lehum}, the case of a single field model with a spurion background seems to adequately respond to the supersymmetric completion. Within this framework, and taking the bosonic sector, Eq. \eqref{Loff} reduces to
\begin{equation}
\begin{aligned}
    \mathcal{L}_{\textrm{off}} &= \frac{1}{2}\phi\Box\phi
    \!+\!\frac{1}{2}\!F^2 \!-\! F\PC{W_{\phi} \!+\! \sigma\mathcal{F}_{\phi}\!}\\
    &-G\PC{\mathcal{F}+\sigma + \sigma\mathcal{F}_{\sigma}}.
\end{aligned}
\end{equation}
The corresponding equation of motion is given by
\begin{equation}
    \partial_{\mu}\partial^{\mu}\phi - F\PC{W_{\phi\phi} + \sigma\mathcal{F}_{\phi\phi}} - G\PC{\mathcal{F}_{\phi} + \sigma\mathcal{F}_{\sigma\phi}} = 0.
\end{equation}
In the static sector, using the relations in Eqs. \eqref{Frel} and \eqref{Grel}, we get that
\begin{equation}
\label{sEoML}
    \begin{aligned}
        \phi^{\prime\prime} &= W_{\phi}W_{\phi\phi} + \sigma\PC{W_{\phi\phi}\mathcal{F}_{\phi} + W_{\phi}\mathcal{F}_{\phi\phi}} + \sigma^2\mathcal{F}_{\phi\phi}\mathcal{F}_{\phi}\\
        &+\eta\sigma^{\prime}\PC{\mathcal{F}_{\phi} + \sigma\mathcal{F}_{\sigma\phi}}.
    \end{aligned}
\end{equation}
This equation reduces to the standard expression $\phi^{\prime\prime} = W_{\phi}W_{\phi\phi}$ when $\mathcal{F}(\Phi,\Sigma)$ is constant, which is expected since this choice decouples the spurion from the matter field. Moreover, by choosing $\mathcal{F}(\Phi,\Sigma) = \Phi$, Eq. \eqref{sEoML} becomes
\begin{equation}
\label{sEoMA}
    \phi^{\prime\prime} = W_{\phi}W_{\phi\phi} + \sigma W_{\phi\phi} + \eta\sigma^{\prime},
\end{equation}
which reproduces the half-BPS soliton-impurity model discussed in Ref. \cite{Adam3}, provided that the scalar potential is defined as $V(\phi) = W_{\phi}^{2}/2$, the impurity is rescaled to $\sqrt{2}\sigma$ and $\eta = -1$. 

In Ref. \cite{Adam3}, the authors considered a kink-form preserving impurity. This corresponds to a particular choice of $\sigma(x)$ for which the original kink or antikink remains a non-BPS solution of the static equation of motion, i.e. $\phi^{\prime\prime} = W_{\phi}W_{\phi\phi} \rightarrow \phi^{\prime} = \pm W_{\phi}$. For Eq. \eqref{sEoMA}, this condition implies
\begin{equation}
    \sigma W_{\phi\phi} + \eta\sigma^{\prime} = 0 \to \eta\sigma^{\prime} \pm \sigma\frac{d}{dx}\ln\abs{\phi^{\prime}} =0.
\end{equation}
This differential equation can be solved analytically, and its solution depends on the sign of $\mp\eta$. One obtains
\begin{equation}
\label{adamimp}
    \sigma(x) = C\PC{\phi^{\prime}}^{\mp\eta},
\end{equation}
where $C$ is a real constant.

However, for the more general static equation \eqref{sEoML}, imposing the kink-form preserving impurity leads to
\begin{equation}
\label{kpr}
    \begin{aligned}
         \sigma\PC{W_{\phi\phi}\mathcal{F}_{\phi} + W_{\phi}\mathcal{F}_{\phi\phi}} &+ \sigma^2\mathcal{F}_{\phi\phi}\mathcal{F}_{\phi}\\
        &+\eta\sigma^{\prime}\PC{\mathcal{F}_{\phi} + \sigma\mathcal{F}_{\sigma\phi}} =0.
    \end{aligned}
\end{equation}
In general, this equation cannot be solved analytically for arbitrary choices of $\mathcal{F}(\Phi,\Sigma)$. A useful simplification is to assume that $\mathcal{F}$ is a function of only the matter field, which results in $\mathcal{F}_{\sigma\phi} = 0$. With this choice and after some algebraic manipulations, one obtains
\begin{equation}
\label{inter}
    \sigma\frac{d}{d\phi}\PC{W_{\phi}\mathcal{F}_{\phi}} + \eta\sigma^{\prime}\mathcal{F}_{\phi} +\frac{\sigma^{2}}{2}\frac{d}{d\phi}\PC{\mathcal{F}_{\phi}^{2}} = 0.
\end{equation}
Multiplying Eq. \eqref{inter} by $\phi^{\prime}$ and using the relation $\phi^{\prime} = \pm W_{\phi}$, yields
\begin{equation}
\label{inter2}
    \sigma\frac{d}{dx}\PC{W_{\phi}\mathcal{F}_{\phi}} \pm \eta\sigma^{\prime}W_{\phi}\mathcal{F}_{\phi} +\frac{\sigma^{2}}{2}\frac{d}{dx}\PC{\mathcal{F}_{\phi}^{2}} = 0.
\end{equation}
The solution of this equation also depends on the sign of $\pm\eta$. Choosing $\pm\eta = 1$, this equation can be written as
\begin{equation}
    \frac{d}{dx}\PC{\sigma_{+} W_{\phi}\mathcal{F}_{\phi}} + \sigma_{+}^2\mathcal{F}_{\phi}\mathcal{F}_{\phi\phi}\phi^{\prime} = 0,
\end{equation}
which can be further manipulated to give
\begin{equation}
\label{sigp}
    \sigma_{+}(x) = \frac{1}{W_{\phi}\mathcal{F}_{\phi}}\PC{\frac{1}{C+\int^{x}\frac{\mathcal{F}_{\phi\phi}}{W_{\phi}^2\mathcal{F}_{\phi}}\frac{d\phi}{d\bar{x}}d\bar{x}}}.
\end{equation}
In this expression, it should be emphasized that all functions are evaluated on the kink or antikink profile $\phi(x)$.

On the other hand, taking $\pm\eta = -1$ in Eq. \eqref{inter2}, gives
\begin{equation}
    \frac{d}{dx}\PC{\frac{W_{\phi}\mathcal{F}_{\phi}}{\sigma_{-}}} = \frac{d}{dx}\PC{-\frac{\mathcal{F}_{\phi}^2}{2}},
\end{equation}
which results in
\begin{equation}
\label{sigm}
    \sigma_{-}(x) = \frac{W_{\phi}\mathcal{F}_{\phi}}{C-\mathcal{F}_{\phi}^2/2}.
\end{equation}
Again, all functions must be evaluated on $\phi(x)$. It should be stressed that the kink-preserving formalism only ensures that the original kink (or antikink) is a non-BPS solution of the equation of motion. Consequently, it does not satisfy the half-BPS equation \eqref{eqBPS} and does not saturate the energy lower bound.

For the specific choice $\mathcal{F}(\Phi,\Sigma) = \Phi$, $\sigma_{+}$ reduces to
\begin{equation}
    \sigma_{+} = \frac{1}{CW_{\phi}} = \frac{C_{1}}{\phi^{\prime}},
\end{equation}
which reproduces Eq. \eqref{adamimp}, when $\mp\eta = -1$. Analogously, the $\sigma_{-}$ becomes
\begin{equation}
    \sigma_{-} = \frac{W_{\phi}}{C - 1/2} = C_{2}\phi^{\prime},
\end{equation}
that also reproduces Eq. \eqref{adamimp}, now for the choice $\mp\eta = 1$.

The analysis of the kink-preserving impurity for the most general case, where the coupling function $\mathcal{F}\PC{\Phi,\Sigma}$ depends on both the matter and spurion fields, is much more involved. Therefore, in these cases, we restrict our discussion to analytical solutions obtained outside the kink-preserving framework.

For the energy density, it is useful to substitute the half-BPS equation in Eq. \eqref{rho}. This gives, for the single field framework
\begin{equation}
\label{rho1f}
    \rho = \PC{W_{\phi} + \sigma\mathcal{F}_{\phi}}^2 + \eta\sigma^{\prime}\PC{\mathcal{F}+\sigma + \sigma\mathcal{F}_{\sigma}}.
\end{equation}
It should be stressed that the second term may not be positive, so the solutions could have regions of negative energy density.

\subsection*{A.1. Examples}

To illustrate the one-field kink-impurity framework previously developed, we consider the $\phi^4$ model, which is defined by
\begin{equation}
    V(\phi) = \frac{1}{2}\PC{1-\phi^2}^2, \quad W(\phi) = \phi-\frac{1}{3}\phi^3.
\end{equation}
Let us now suppose that the coupling function depends only on the matter field. The choice of this function that results in an analytical localized $\sigma_{+}$ and a well-defined half-BPS equation is not straightforward, since the kink-preserving impurity $\sigma_{+}$ depends on a nontrivial integral expression. A convenient possibility is to take
\begin{equation}
\label{f1}
    \mathcal{F}(\phi) = \phi\,{}_2F_1\!\PC{-\frac{1}{3},\frac{1}{2};\frac{3}{2};\phi^2},
\end{equation}
where ${}_2F_1$ is the Gaussian hypergeometric function, which is real and finite for $\abs{\phi}\leq1$. This choice implies that
\begin{equation}
    \mathcal{F}_{\phi} = (1-\phi^2)^{1/3}.
\end{equation}

\begin{figure}[!ht]
    \centering
\includegraphics[width=0.9\columnwidth,height=5.0cm]{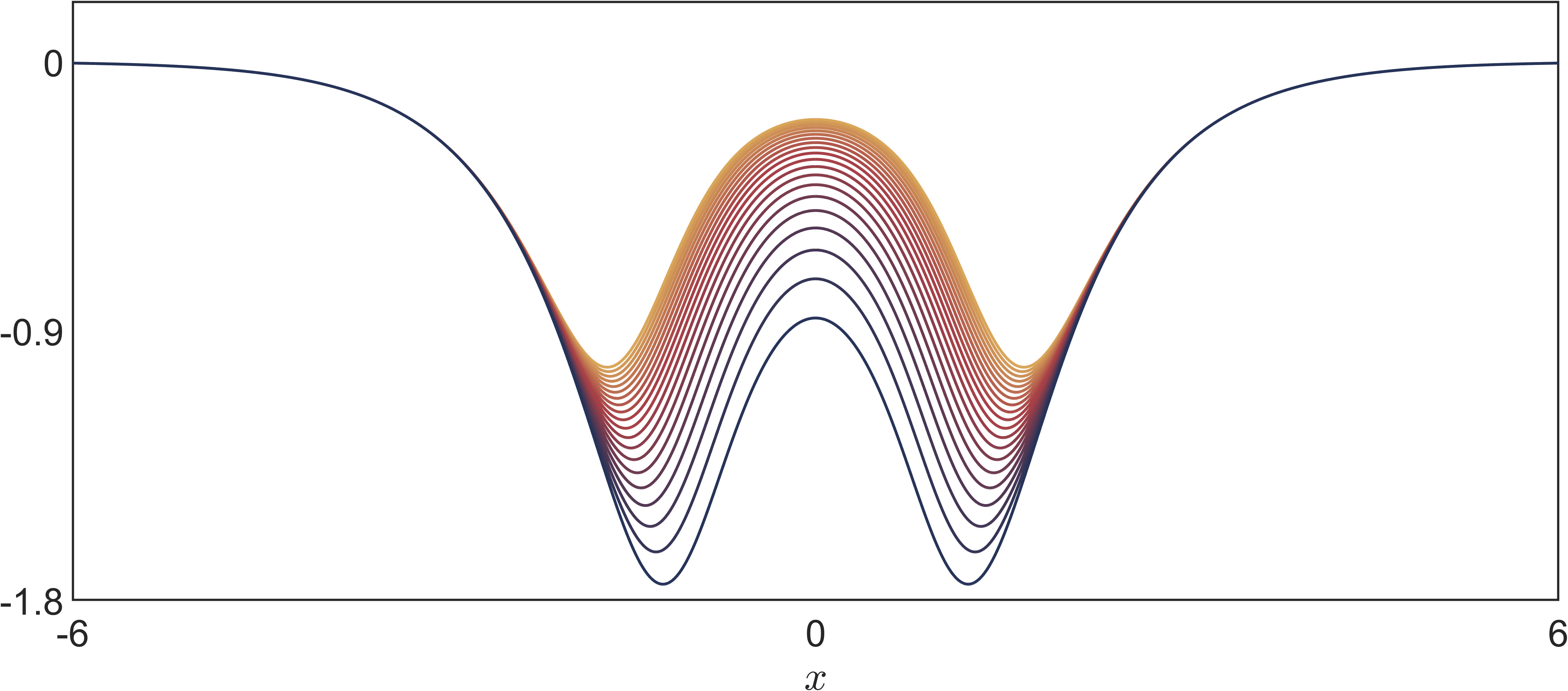}
    \caption{Kink-preserving impurity $\sigma_{+}(x)$ given by Eq. \eqref{sig1}. The parameter $C$ is varied from $-5$ to $-1$, with the colors changing from lighter to darker.}
    \label{fig1}
\end{figure}

To obtain an explicit kink-preserving impurity, we choose the non-BPS solution as the standard $\phi^4$ kink $\phi(x) = \tanh(x)$, where its center is at $x = 0$ for simplicity. Therefore, Eq. \eqref{sigp} leads to the analytical impurity
\begin{equation}
\label{sig1}
    \sigma_{+}(x) = \frac{\cosh^{8/3}(x)}{C-\frac{1}{6}\cosh^{4}(x)}.
\end{equation}
This impurity is localized, since $\sigma_{+}(x)$ behaves asymptotically as $\sigma_{+}\propto\sech^{4/3}(x)$. It is also nonsingular for $C<\frac{1}{6}$. The $\sigma_{+}$ behavior and its dependence on the parameter $C$ are depicted in Fig. \ref{fig1}. Note that this impurity has an internal structure and is centered at $x=0$.

\begin{figure}[!ht]
    \centering
\includegraphics[width=0.9\columnwidth,height=5.0cm]{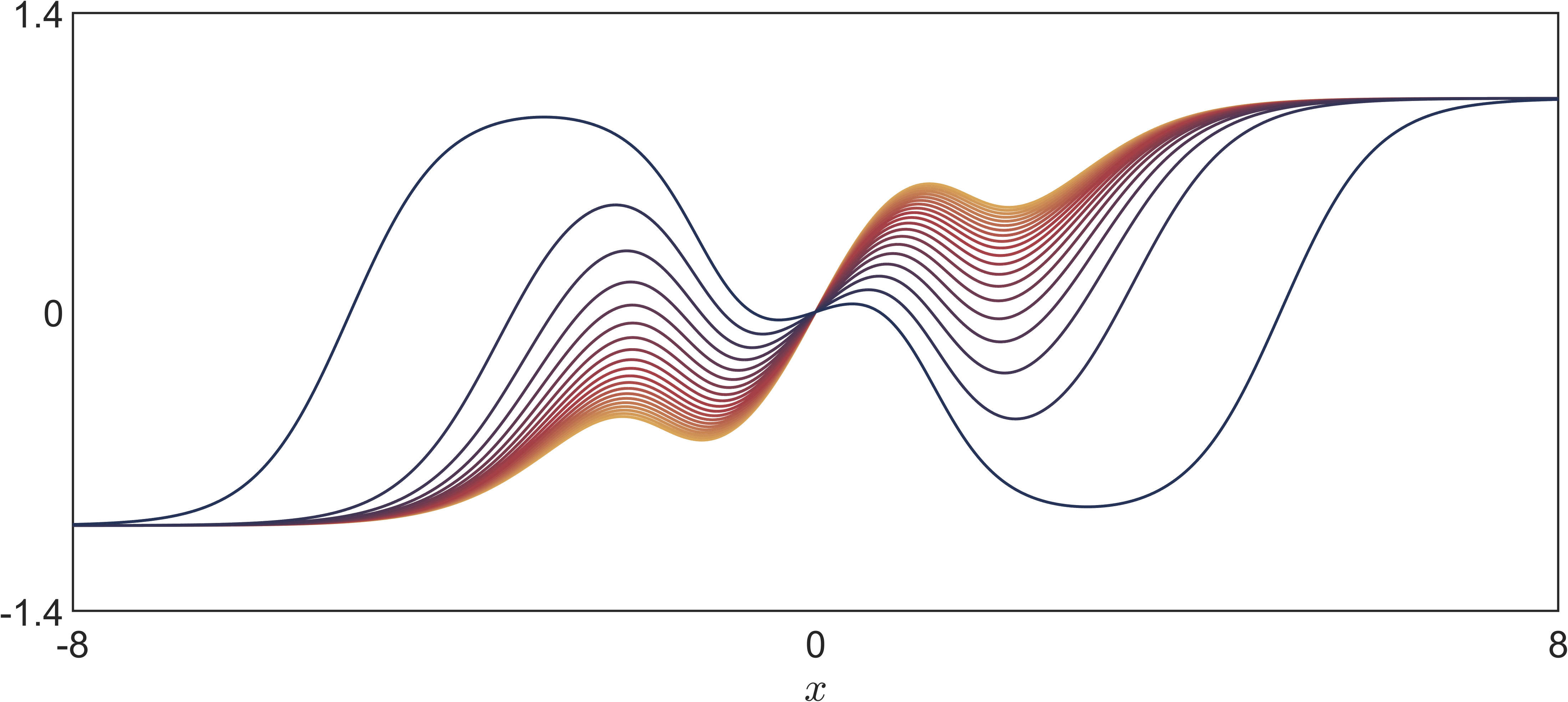}
    \caption{Half-BPS kink solutions obtained by numerically solving Eq. \eqref{fo1} for different values of the parameter $C$ in the interval $-5\leq C \leq -1$, with the colors changing from lighter to darker.}
    \label{fig2}
\end{figure}

The corresponding half-BPS equation is
\begin{equation}
\label{fo1}
    \phi' = 1-\phi^2 + \frac{(1-\phi^2)^{1/3}\cosh^{8/3}(x)}{C-\frac{1}{6}\cosh^{4}(x)}.
\end{equation}
Thus, this choice of $\mathcal{F}(\phi)$ also provides a half-BPS equation without poles, for $C<1/6$. This first-order differential equation was solved numerically with the initial condition $\phi(0) = 0$, for different values of $C$. The resulting solutions are shown in Fig. \ref{fig2}. It should be noted that the internal structure of the impurity is inherited by the kink profile, while the reflection symmetry of $\sigma_{+}(x)$ about $x=0$ is not preserved by the solution because of the kink boundary conditions.

\begin{figure}[!ht]
    \centering
\includegraphics[width=0.9\columnwidth,height=5.0cm]{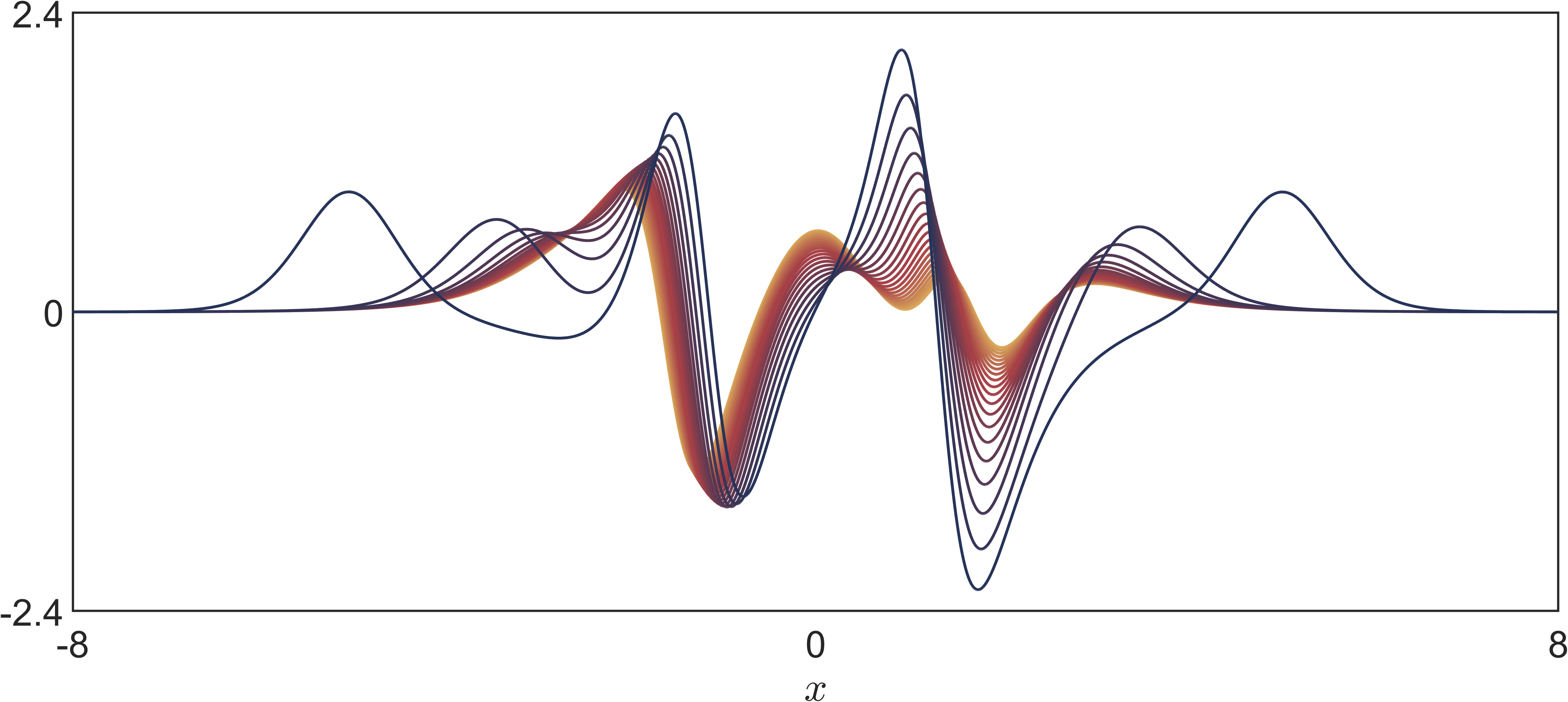}
    \caption{Energy density $\rho(x)$ given by Eq. \eqref{rho1}. The parameter $C$ is varied from $-5$ to $-1$, with the colors changing from lighter to darker.}
    \label{fig3}
\end{figure}

The energy density defined in Eq. \eqref{rho1f} acquires, for this model, the form
\begin{equation}
\label{rho1}
\begin{aligned}
    \rho(x) &= \PC{1-\phi^2 + \frac{\cosh^{8/3}(x)}{C-\frac{1}{6}\cosh^{4}(x)}\PC{1-\phi^2}^{1/3}}^2\\
    &+ \PC{\frac{8\cosh^{5/3}(x)\sinh(x)\PC{12C + \cosh^4(x)}}{\PC{6C - \cosh^4(x)}^2}}\\
    &\times\PC{\phi\,{}_2F_1\!\PC{-\frac{1}{3},\frac{1}{2};\frac{3}{2};\phi^2} + \frac{\cosh^{8/3}(x)}{C-\frac{1}{6}\cosh^{4}(x)}}.
\end{aligned}
\end{equation}
The profile of $\rho(x)$ is shown in Fig. \ref{fig3} for different values of $C$. As expected, the solutions also provide regions with negative energy densities, a property enhanced by its non-monotonic profile.

\begin{figure}[!ht]
    \centering
\includegraphics[width=0.9\columnwidth,height=5.0cm]{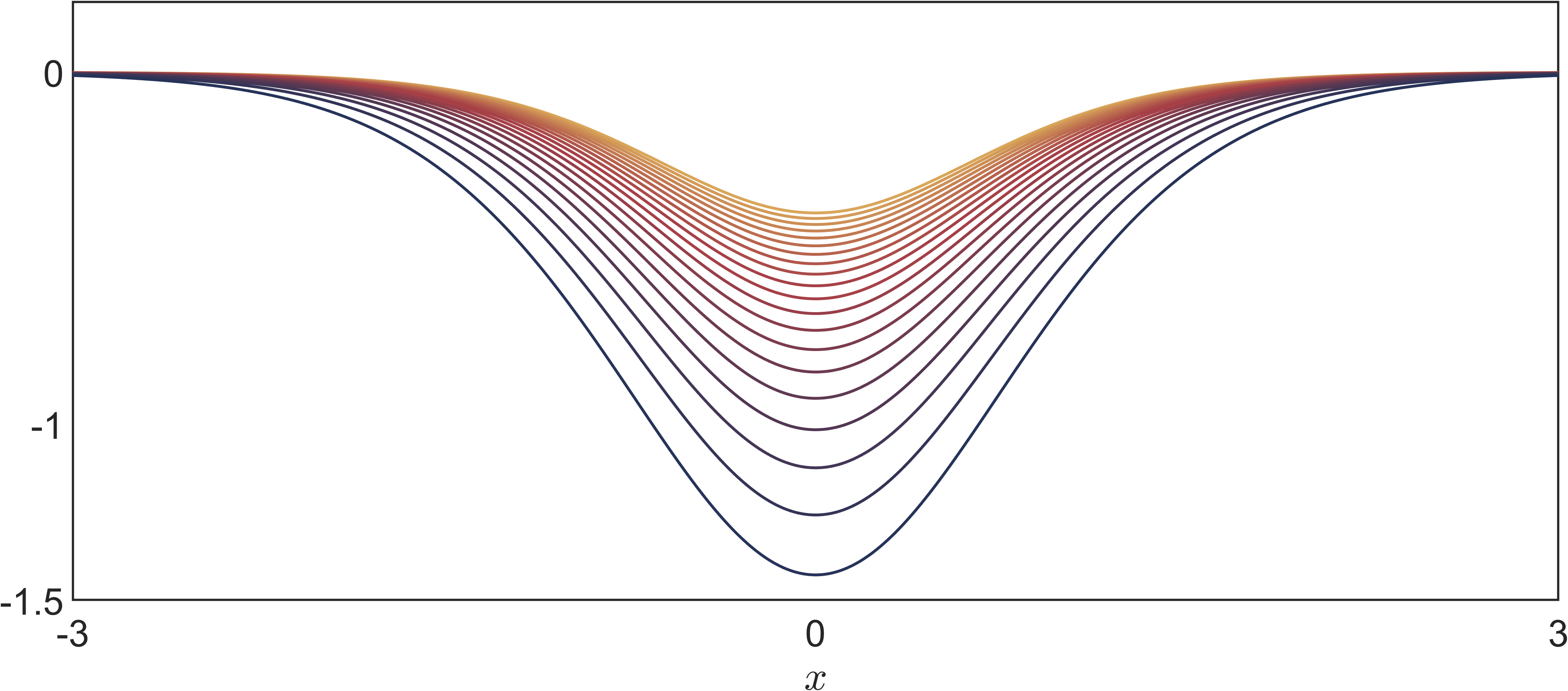}
\includegraphics[width=0.9\columnwidth,height=5.0cm]{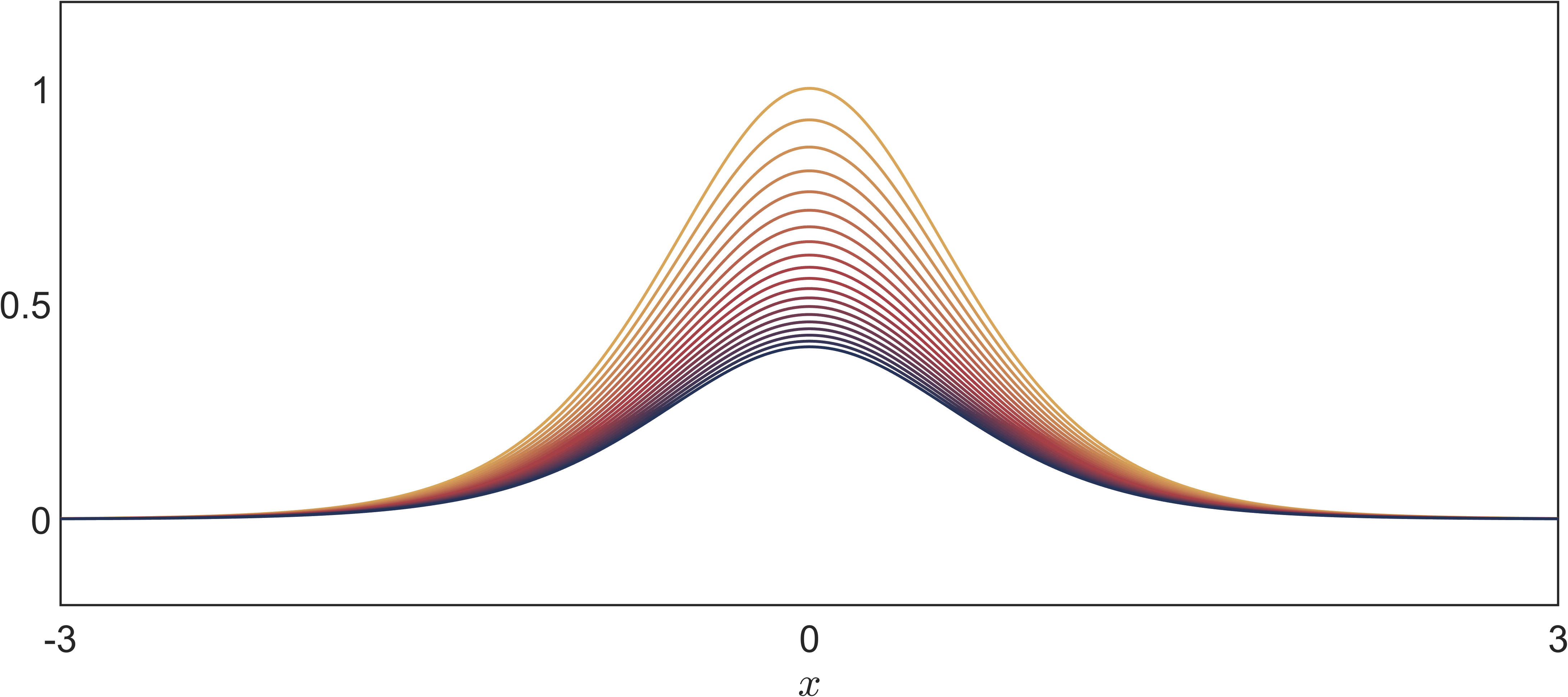}
    \caption{Kink-preserving impurity $\sigma_{-}$ given by Eq. \eqref{sig2}. In the upper panel, the parameter $C$ is varied in the interval $-2 \leq C \leq -0.2$, while in the lower panel it is varied in the interval $1.5 \leq C \leq 3$. In both cases, the colors change from lighter to darker.}
    \label{fig4}
\end{figure}

This choice of $\mathcal{F}(\phi)$ also works to define a localized and well-behaved $\sigma_{-}$ kink-preserving impurity. Substituting the coupling function $\mathcal{F}(\phi)$ defined in Eq. \eqref{f1} into Eq. \eqref{sigm} and choosing the non-BPS solution as the standard $\phi^4$ kink, one obtains the analytical impurity
\begin{equation}
\label{sig2}
    \sigma_{-}(x) = \frac{\sech^{8/3}(x)}{C-\frac{1}{2}\sech^{4/3}(x)},
\end{equation}
which is localized and has no poles for $C\leq0$ or $C>1/2$. The $\sigma_{-}$ behavior and its dependence on the parameter $C$ are shown in Fig. \ref{fig4}.

It should be noted that, unlike the $\sigma_{+}$ case, for $\sigma_{-}$ there is no choice of the parameter $C$ that leads to the formation of internal structures on the impurity, since within the range of $C$ considered, $\sigma_{-}$ has only one critical point at $x=0$.

\begin{figure}[!ht]
    \centering
\includegraphics[width=0.9\columnwidth,height=5.0cm]{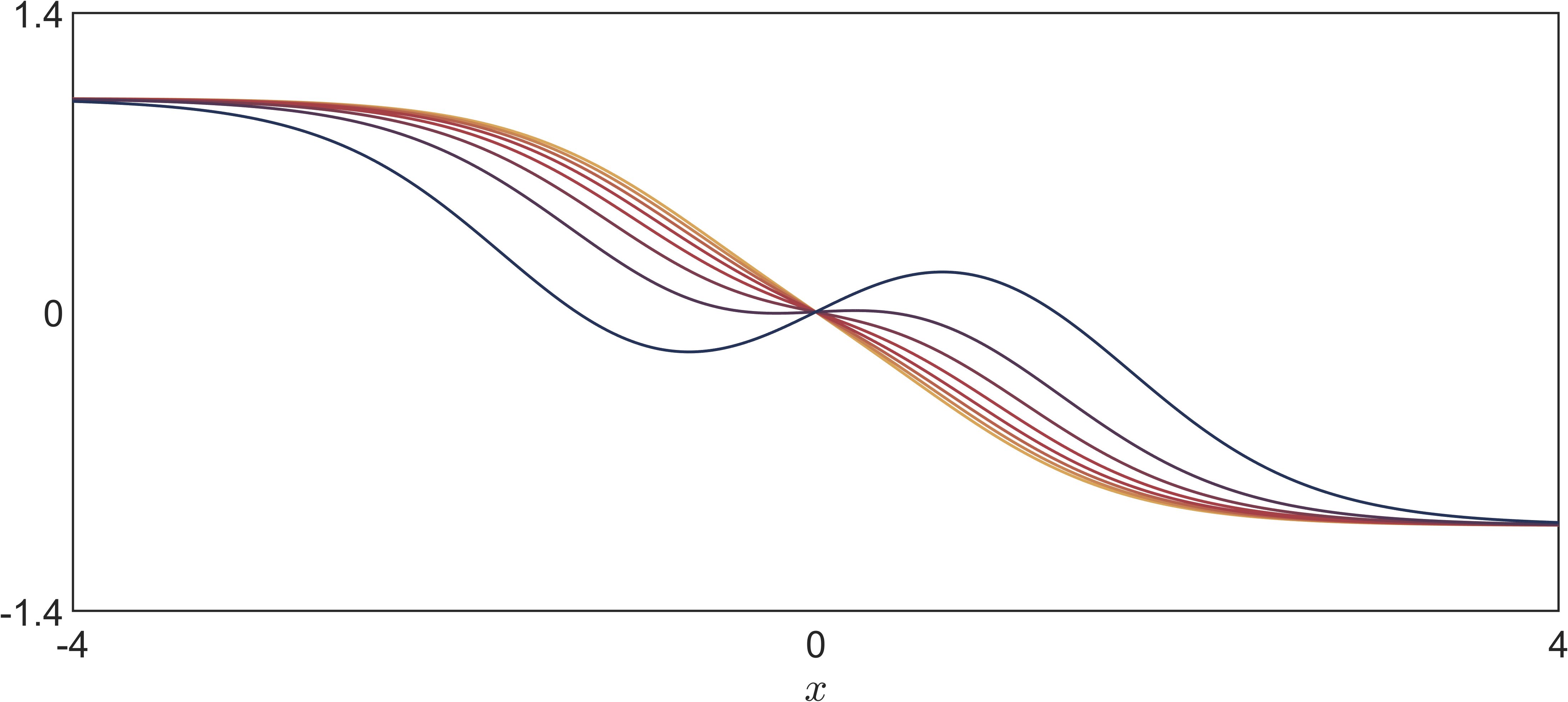}
\includegraphics[width=0.9\columnwidth,height=5.0cm]{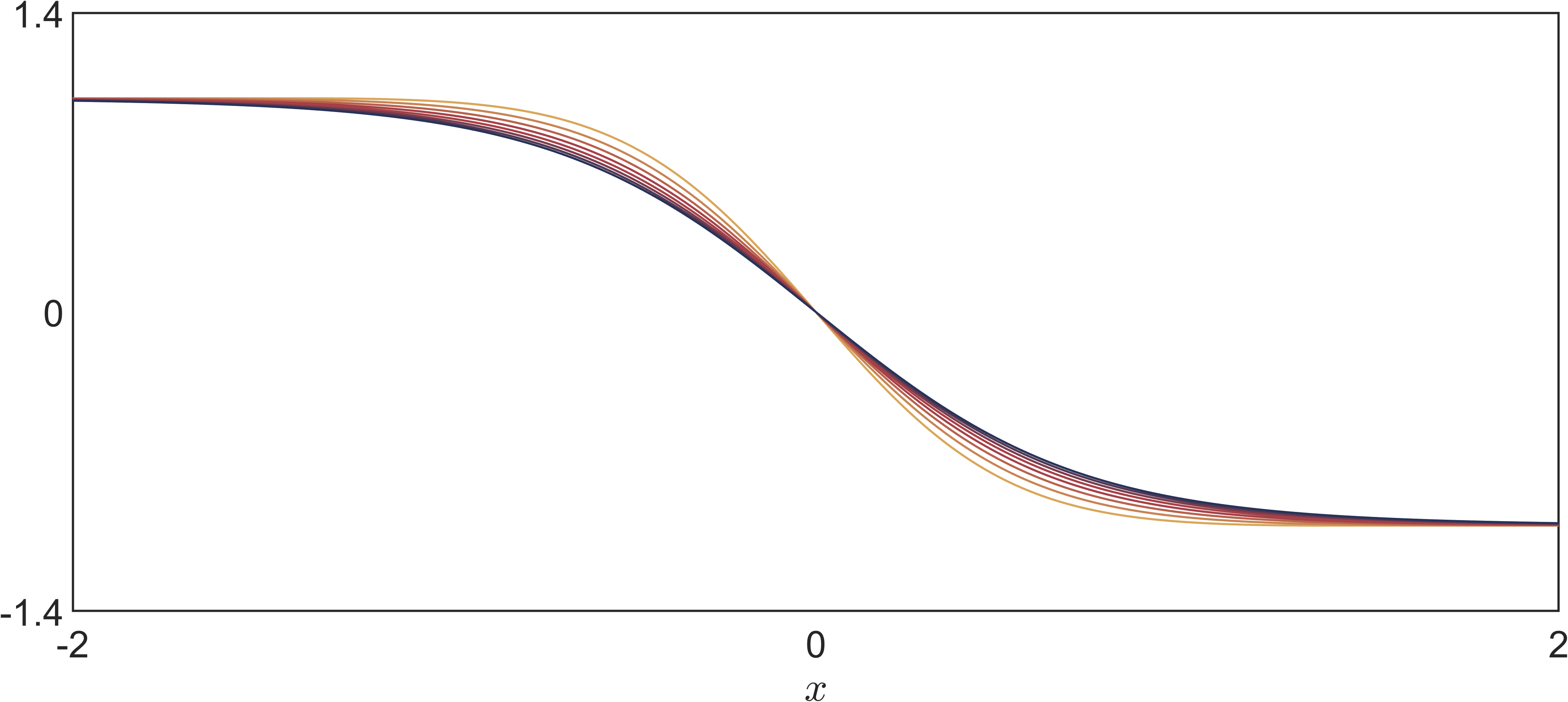}
    \caption{Half-BPS kink solutions obtained by numerically solving Eq. \eqref{fo2} for different values of the parameter $C$. The parameter $C$ is varied from $-2$ to $-0.2$ in the upper panel and from $1.5$ to $3$ on the bottom panel. In both cases, the colors change from lighter to darker.}
    \label{fig5}
\end{figure}

The corresponding half-BPS equation becomes
\begin{equation}
\label{fo2}
    \phi' = -\PC{1-\phi^2 + \frac{(1-\phi^2)^{1/3}\sech^{8/3}(x)}{C-\frac{1}{2}\sech^{4/3}(x)}}.
\end{equation}
This equation was also solved numerically, with the initial condition of $\phi(0) = 0$. The solutions are displayed in Fig. \ref{fig5}, where the dependence of the profile on the parameter $C$ is shown. For negative values of the parameter $C$, the solution develops an internal structure. This behavior does not come from the impurity profile itself, but from the nonlinear coupling between the scalar field and the impurity in the half-BPS equation \eqref{fo2}.

\begin{figure}[!ht]
    \centering
\includegraphics[width=0.9\columnwidth,height=5.0cm]{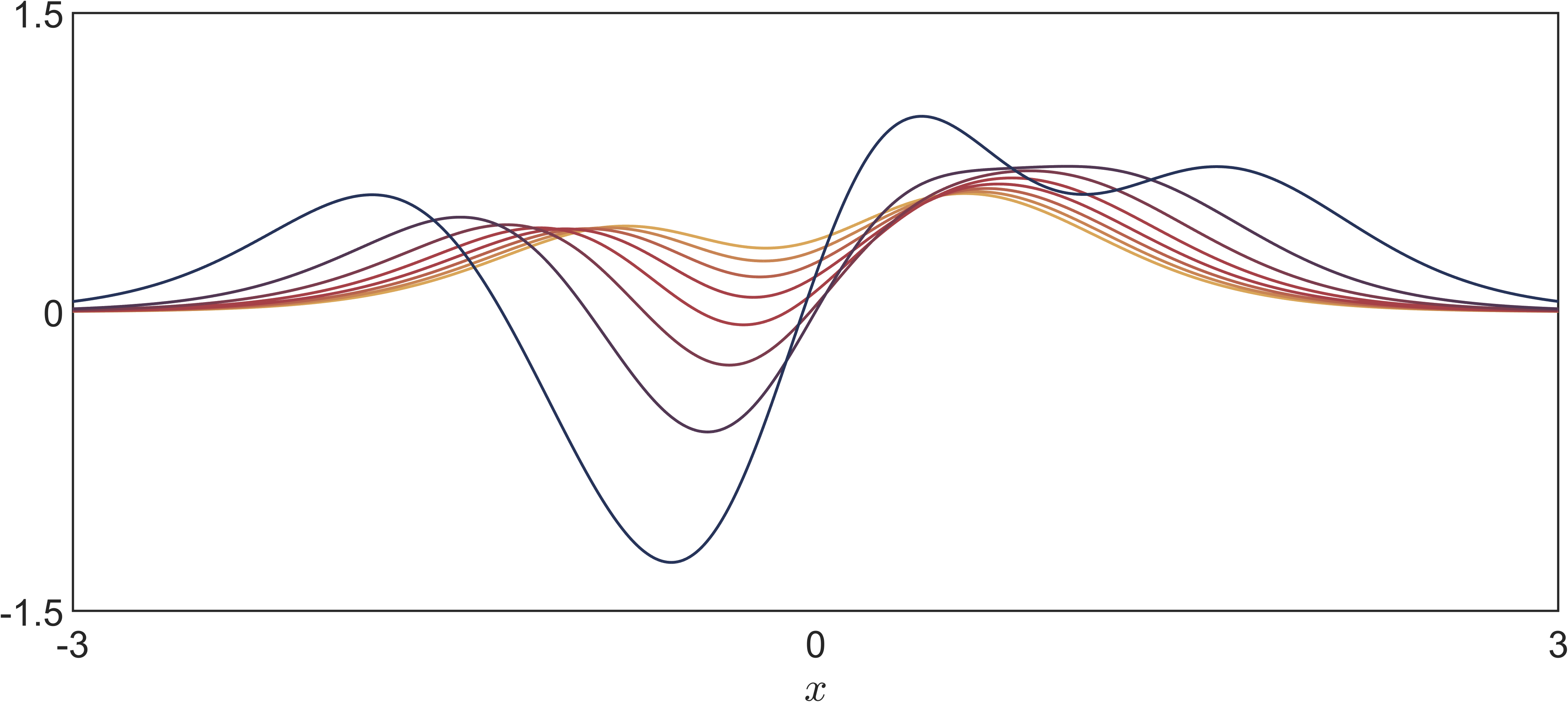}
\includegraphics[width=0.9\columnwidth,height=5.0cm]{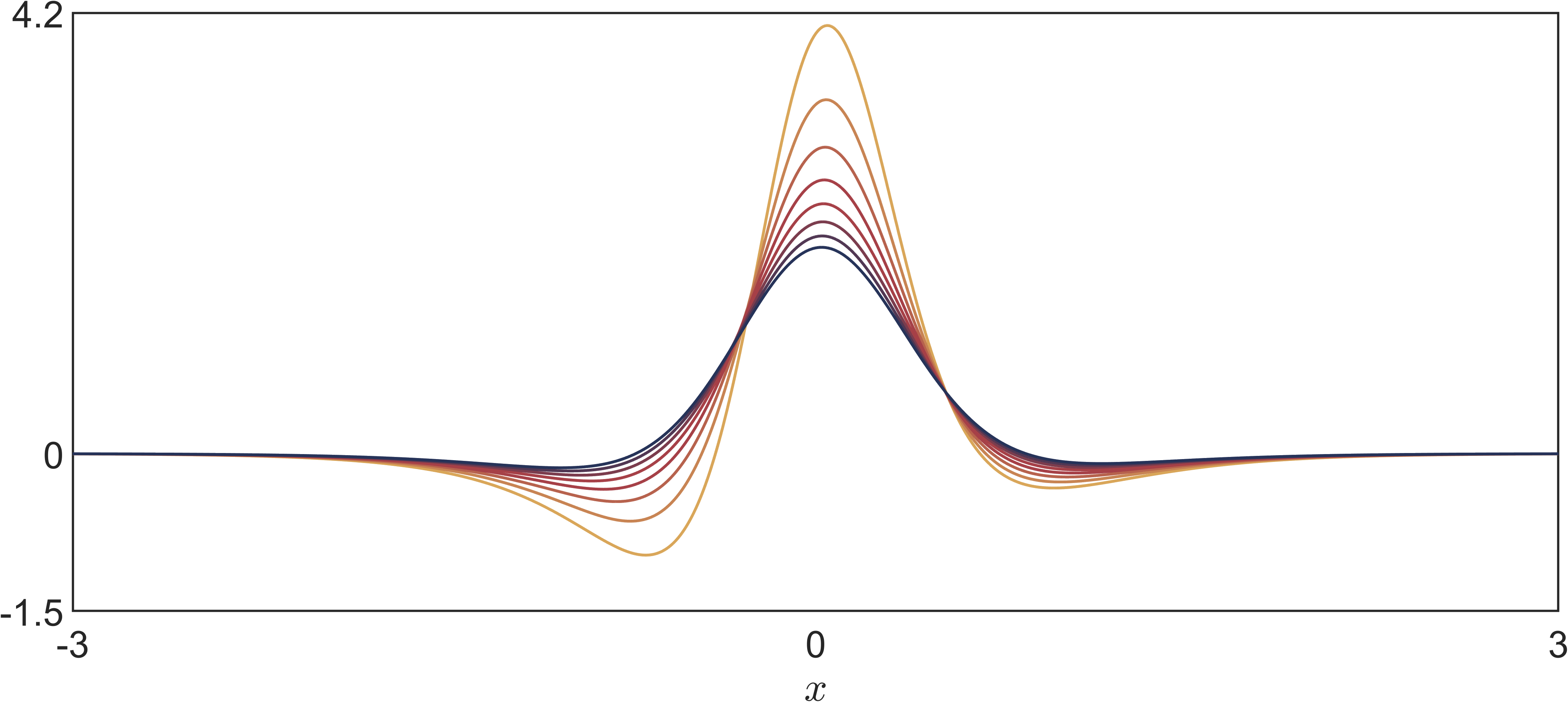}
    \caption{Energy density $\rho(x)$ given by Eq. \eqref{rho2}. In the upper panel, the parameter $C$ is varied in the interval $-2 \leq C \leq -0.2$, while in the lower panel it is varied in the interval $1.5 \leq C \leq 3$. In both panels, the colors change from lighter to darker.}
    \label{fig6}
\end{figure}

For this model, the energy density \eqref{rho1f} results in
\begin{equation}
\label{rho2}
\begin{aligned}
    \rho(x) &= \PC{1-\phi^2 + \frac{\sech^{8/3}(x)}{C-\frac{1}{2}\sech^{4/3}(x)}\PC{1-\phi^2}^{1/3}}^2\\
    &- \PC{\frac{8\tanh(x)\PC{\sech^4(x) -4C\sech^{8/3}(x)}}{3\PC{2C - \sech^{4/3}(x)}^2}}\\
    &\times\PC{\phi\,{}_2F_1\!\PC{-\frac{1}{3},\frac{1}{2};\frac{3}{2};\phi^2} + \frac{\sech^{8/3}(x)}{C-\frac{1}{2}\sech^{4/3}(x)}}.
\end{aligned}
\end{equation}
The profile of the energy density is shown in Fig. \ref{fig6}, exhibiting regions with negative values even when the solution profiles remain monotonic. However, when they are non-monotonic, the energy density acquires a richer internal structure.

It is worth emphasizing that finding a single definition of the coupling function $\mathcal{F}(\phi)$ capable of producing both $\sigma_{+}$ and $\sigma_{-}$ well-defined and localized, while also leading to regular half-BPS equations, is not typical. In the model discussed in Ref. \cite{Adam3}, for instance, this is impossible. This happens because, as we have already shown, $\sigma_{+}\propto1/\phi^{\prime}$ and $\sigma_{-}\propto\phi^{\prime}$ for  $\mathcal{F}(\phi) = \phi$; thus, both impurities cannot be localized simultaneously.

\begin{figure}[!ht]
    \centering
\includegraphics[width=0.9\columnwidth,height=5.0cm]{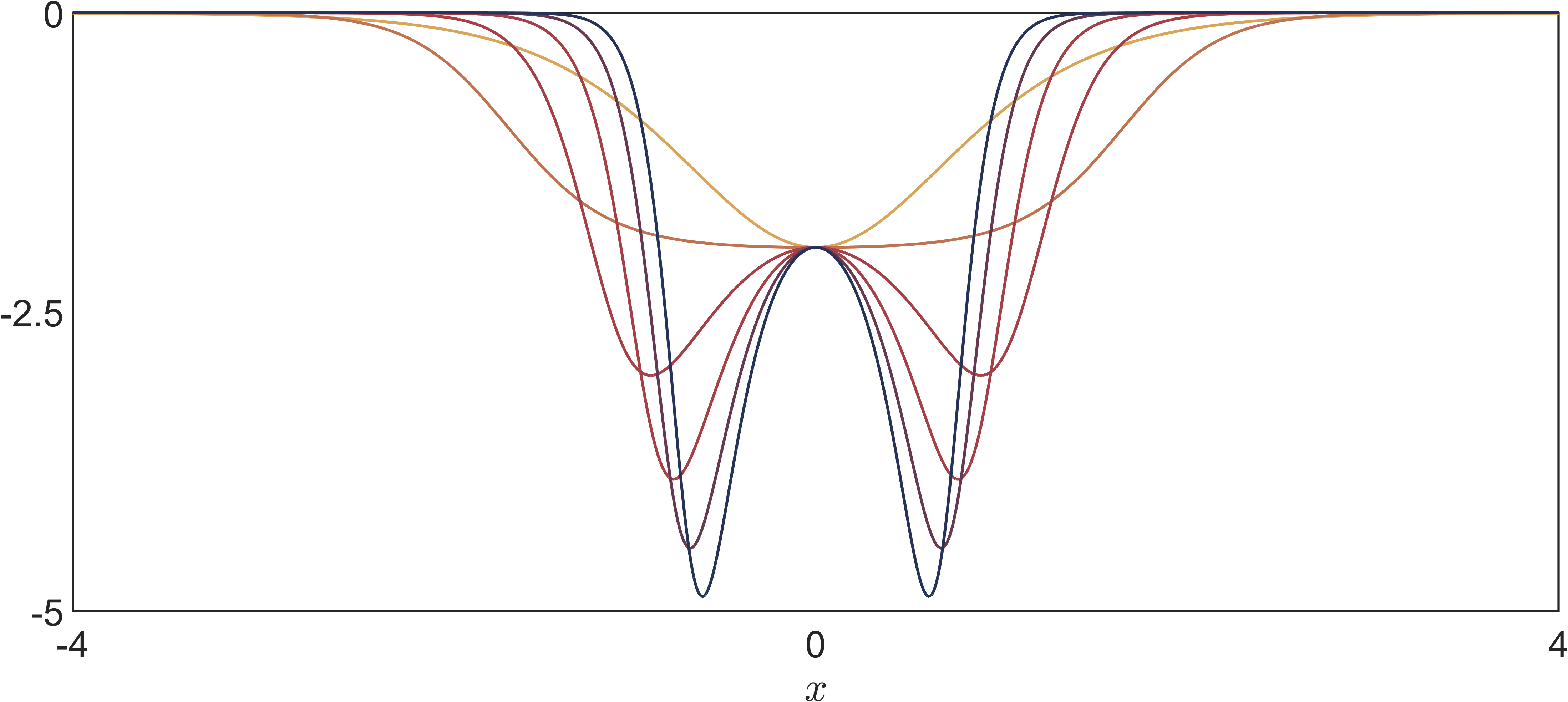}
    \caption{Kink-preserving impurity $\sigma_{-}(x)$ given by Eq. \eqref{sig3} for $C = -0.01$. The parameter $n$ goes from $0$ to $5$, and the colors change from lighter to darker.}
    \label{fig7}
\end{figure}

This obstacle also persists for more general choices of the coupling function. A simple example is obtained by taking $\mathcal{F}_{\phi} = \PC{1-\phi^2}^{n}$, with $n$ a nonnegative integer. Particularly, by taking $n=0$ the model discussed in \cite{Adam3} is recovered. However, in the general case, Eq. \eqref{sigp} leads to
\begin{equation}
    \sigma_{+}(x) = \frac{\cosh^{2(n+1)}(x)}{C-\frac{n}{2}\cosh^4(x)},
\end{equation}
which is not localized for any choice of $n$. However, following Eq. \eqref{sigm}
\begin{equation}
\label{sig3}
    \sigma_{-}(x) = \frac{\sech^{2(n+1)}(x)}{C-\frac{1}{2}\sech^{4n}(x)}.
\end{equation}
To guarantee that this function is nonsingular and localized for any choice of $n$, we must consider $C>1/2$ or $C<0$. When $C$ is negative and close to zero, an internal structure becomes evident in the impurity profile. Otherwise, it behaves as a localized lump. So, in  Fig. \ref{fig7}, we have plotted the impurity profile only for $C = -0.01$ and several values of $n$, to enhance the formation of those structures.

\begin{figure}[!ht]
    \centering
\includegraphics[width=0.9\columnwidth,height=5.0cm]{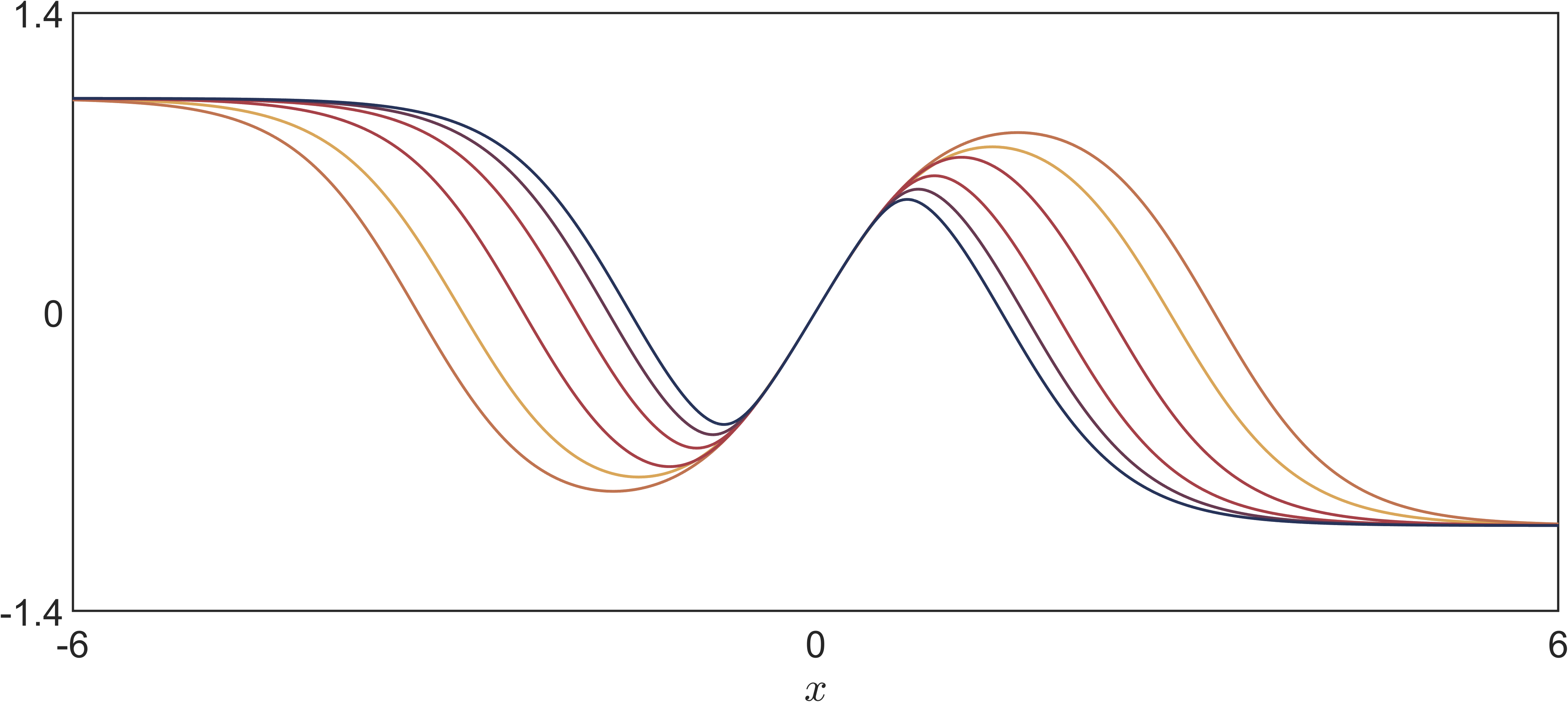}
    \caption{Half-BPS kink solutions obtained by numerically solving Eq. \eqref{fo3} for $C = -0.01$. The parameter $n$ goes from $0$ to $5$, and the colors change from lighter to darker.}
    \label{fig8}
\end{figure}

The corresponding half-BPS equation becomes
\begin{equation}
\label{fo3}
    \phi' = -\PC{1-\phi^2 + \PC{1-\phi^2}^{n}\frac{\sech^{2(n+1)}(x)}{C-\frac{1}{2}\sech^{4n}(x)}}.
\end{equation}
This equation was also solved numerically, with the initial condition $\phi(0) = 0$. The result is plotted in Fig. \ref{fig8}, where the dependence of the profile on the parameter $n$ is displayed.

Thus far, we have proposed impurity profiles that preserve the standard kink solution as a non-BPS solution of the system. However, this is not mandatory. Since the impurity is treated as a background field, one may choose any localized and sufficiently well-behaved profile. Alternatively, instead of requiring the impurity to preserve the standard kink solution, one may choose the coupling function $\mathcal{F}$ in such a way that leads to analytic half-BPS solutions. A useful choice is to consider $\mathcal{F} = W(\phi)\Gamma(\sigma)$, where $\Gamma(\sigma)$ is a general function of the impurity. The half-BPS equation \eqref{eqBPS} reduces to
\begin{equation}
\label{foa}
    \phi^{\prime} = \eta W_{\phi}\PR{1 + \sigma(x)\Gamma(\sigma(x))}.
\end{equation}
The solution can be written as $\phi(x) = \phi_0(\xi(x))$, where $\phi_0(x)$ is the standard kink profile and
\begin{equation}
\label{xi}
    \xi(x) = \int\PC{1 + \sigma(x)\Gamma(\sigma(x))}dx.
\end{equation}

We note that if $\xi(x)$ is strictly monotonic and has a nonvanishing derivative, it defines a global coordinate and provides a one-to-one reparametrization of the spatial variable. In this case, the first-order equation can be rewritten in the standard form
\begin{equation}
    \frac{d\phi_{0}}{d\xi} = \eta W_{\phi}.
\end{equation}
However, if $\xi(x)$ is non-monotonic, the correspondence between $\xi$ and $x$ is not one-to-one, meaning that $\xi(x)$ is not globally invertible. Nevertheless, it may still be interpreted as a local coordinate on each interval over which $\xi^{\prime}(x)$ is nonvanishing and has a definite sign, requiring continuity over the boundary of these intervals. At isolated points where $\xi^{\prime}(x) = 0$, the coordinate transformation becomes singular, although the solution remains well defined by continuity. In this sense, even when $\xi(x)$ cannot be interpreted as a global coordinate, the composition $\phi(x) = \phi_0(\xi(x))$ remains well defined and provides a solution of the original first-order equation \eqref{foa}. A similar reparameterization was previously introduced in Ref. \cite{NewRef} for a single-field impurity model. Here, this mechanism is applied to a broader class of single-field impurity couplings and below extended to multifield models.

\begin{figure}[!ht]
    \centering
\includegraphics[width=0.9\columnwidth,height=5cm]{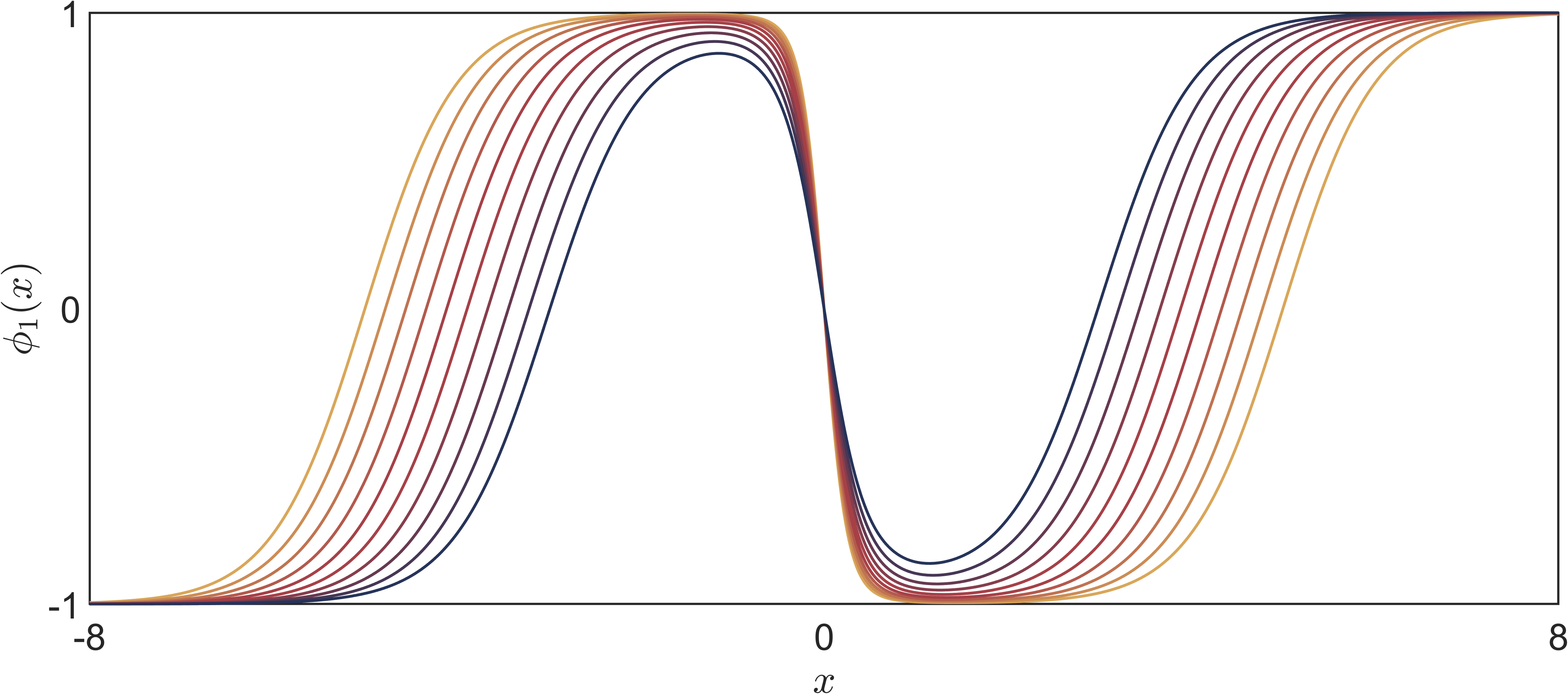}
\includegraphics[width=0.9\columnwidth,height=5cm]{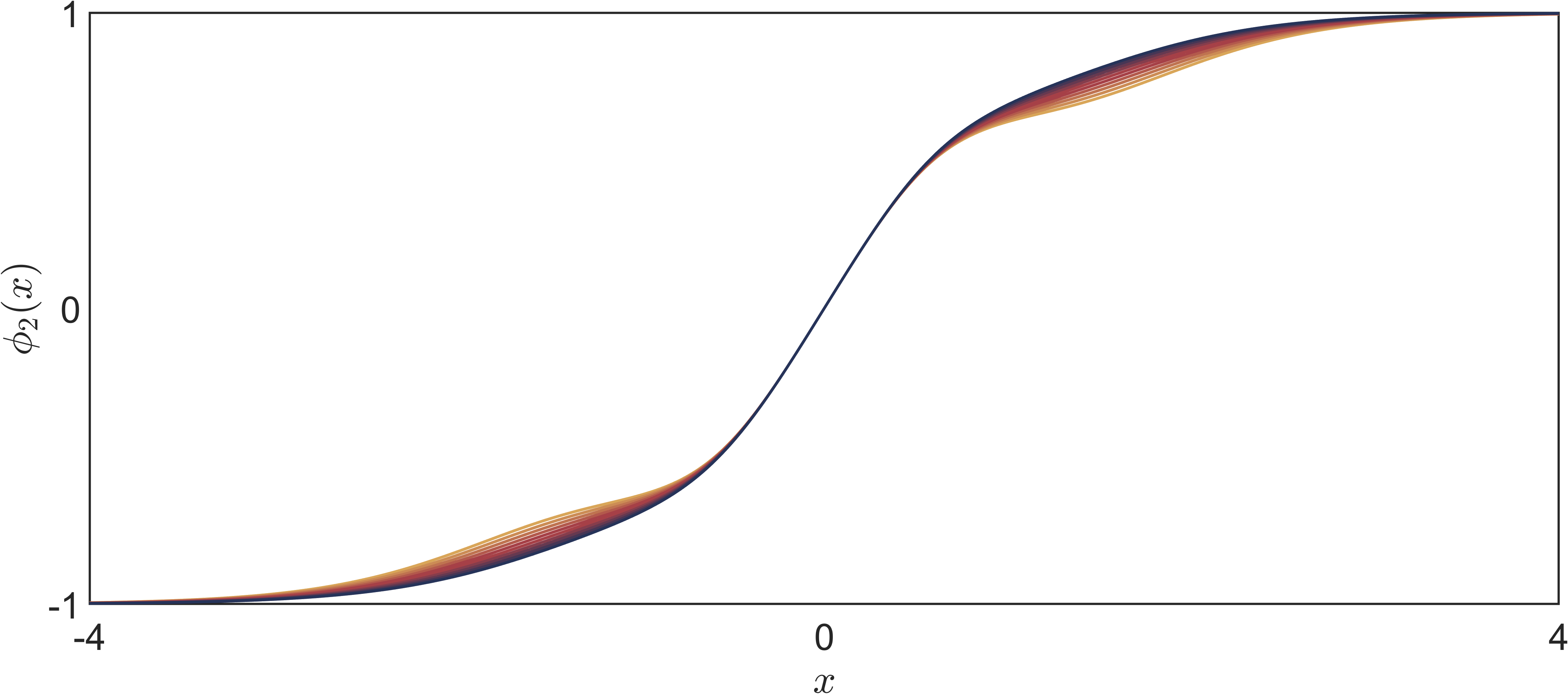}
    \caption{Half-BPS kink solutions $\phi_{1}(x) = \tanh(\xi_{1}(x))$ and $\phi_{2}(x) = \tanh(\xi_{2}(x))$, for $\xi_{1}(x)$ and $\xi_{2}(x)$ defined in Eq. \eqref{xii}. Here, the parameter $\alpha$ is varied from $-5$ to $-3$, with the colors changing from lighter to darker.}
    \label{fig9}
\end{figure}

 For the $\phi^4$ model and $\eta=1$, the solution becomes $\phi(x) = \tanh(\xi(x))$. To illustrate this construction, we consider a simple localized impurity profile, $\sigma(x) = \sech^2(x)$, and two choices for the $\Gamma(\sigma)$ function
 \begin{subequations}
 \begin{align}
    \Gamma_{1}(\sigma(x)) &= \alpha,\\
    \Gamma_{2}(\sigma(x)) &= \alpha\PC{1-\sigma(x)}^2.
 \end{align}
 \end{subequations}
 The parameter $\alpha\in\mathbb{R}$ controls the strength and sign of the coupling between the impurity and the scalar field $\phi$, with $\alpha=0$ corresponding to the decoupled scenario. These choices lead to the functions
\begin{subequations}
\label{xii}
    \begin{align}
        \xi_{1}(x) &= x + \alpha\tanh(x),\\
        \xi_{2}(x) &= x + \alpha\frac{\tanh^5(x)}{5},
    \end{align}
\end{subequations}
where the integration constants were set to zero for simplicity. The corresponding half-BPS solutions $\phi_{i}(x) = \tanh(\xi_{i}(x))$ are plotted in Fig. \ref{fig9} for different values of $\alpha$. Since $\tanh(\xi)$ is a strictly increasing function of $\xi$, the monotonicity of the solutions is determined by the function $\xi_{i}(x)$. For the first choice, $\xi_{1}^{\prime}(x) = 1+\alpha\sech^2(x)$ and the solution becomes non-monotonic for $\alpha<-1$. For the second choice, $\xi_{2}^{\prime}(x) = 1 + \alpha\sech^2(x)\tanh^4(x)$ leads to non-monotonic solutions for $\alpha<-27/4$.

For $\mathcal{F} = W(\phi)\Gamma(\sigma)$, the energy density \eqref{rho1f} reduces to
\begin{equation}
\begin{aligned}
    \rho(x) &= W_{\phi}^2\PC{1+\sigma\Gamma}^2 + \eta\sigma^{\prime}\PC{W\Gamma + \sigma + \sigma W\Gamma_{\sigma}}\\
    &=\eta\frac{d}{dx}\PC{W(\phi(x))\xi^{\prime}(x) + \frac{1}{2}\sigma^{2}(x)}.
\end{aligned}
\end{equation}
Since both the solutions and the impurity profiles are analytical, the energy density can also be expressed analytically. Taking the solution $\phi_{1}$, and the choice $\Gamma_{1}$, the energy density gives
\begin{equation}
\label{rho4}
\begin{aligned}
    \rho_{1}\!(\!x\!) \!=\! \frac{d}{dx}\!\bigg(\!\!\!\PC{\!\tanh(\xi_{1}\!(\!x\!)) \!-\! \frac{\tanh^3\!(\xi_{1}\!(\!x\!))\!}{3}}\!&\PC{\!1\!+\!\alpha\sech^2(\!x\!)\!}\\
    & + \frac{1}{2}\sech^4(\!x\!)\!\bigg).
\end{aligned}
\end{equation}
However, taking $\phi_{2}$ and $\Gamma_{2}$, the energy density gives
\begin{equation}
\label{rho5}
\begin{aligned}
    \rho_{2}\!(\!x\!) \!=\! \frac{d}{dx}\!\bigg(&\PC{\!\tanh(\xi_{2}\!(\!x\!)) \!-\! \frac{\tanh^3\!(\xi_{2}\!(\!x\!))\!}{3}}\\
    &\times\PC{\!1\!+\!\alpha\sech^2(\!x\!)\tanh^4(\!x\!)\!}\\
    & + \frac{1}{2}\sech^4(\!x\!)\!\bigg).
\end{aligned}
\end{equation}

\begin{figure}[!ht]
    \centering
\includegraphics[width=0.9\columnwidth,height=5.0cm]{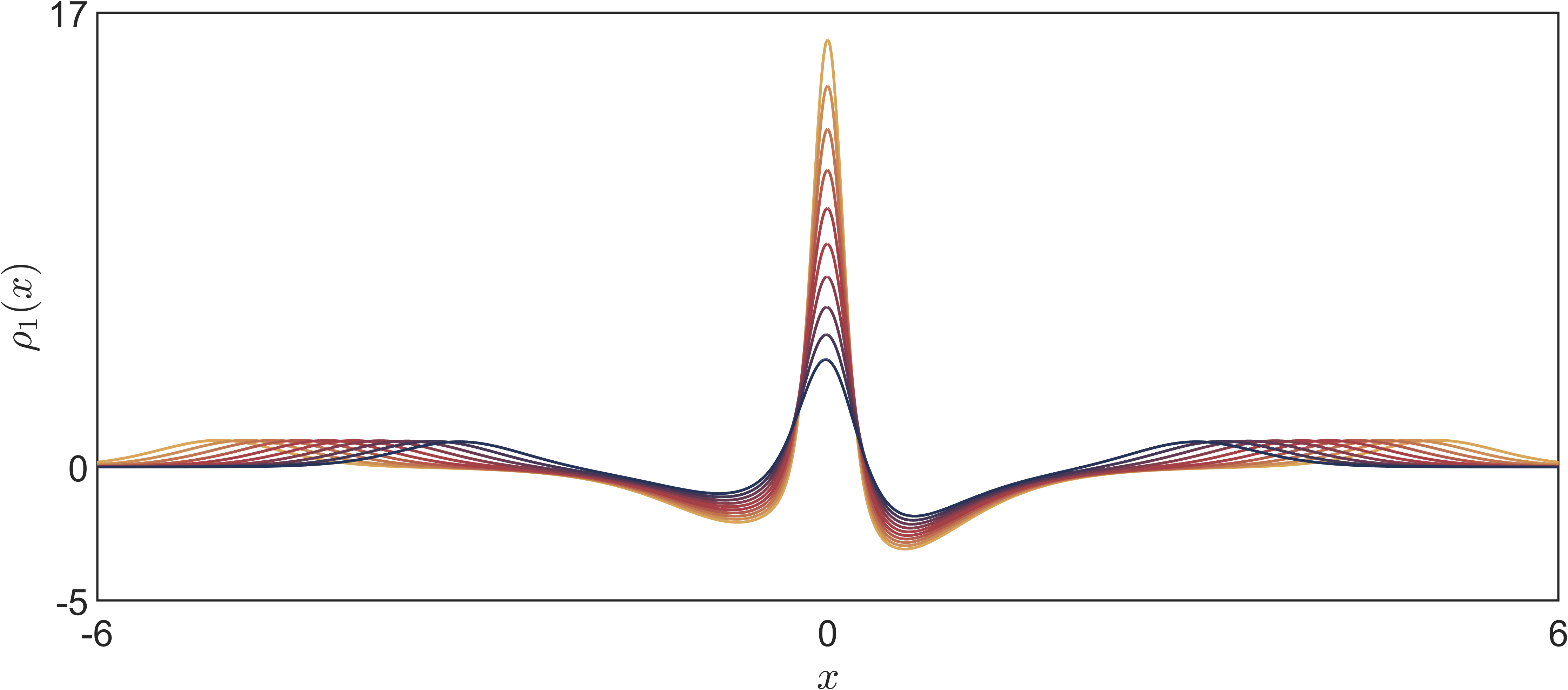}
\includegraphics[width=0.9\columnwidth,height=5.0cm]{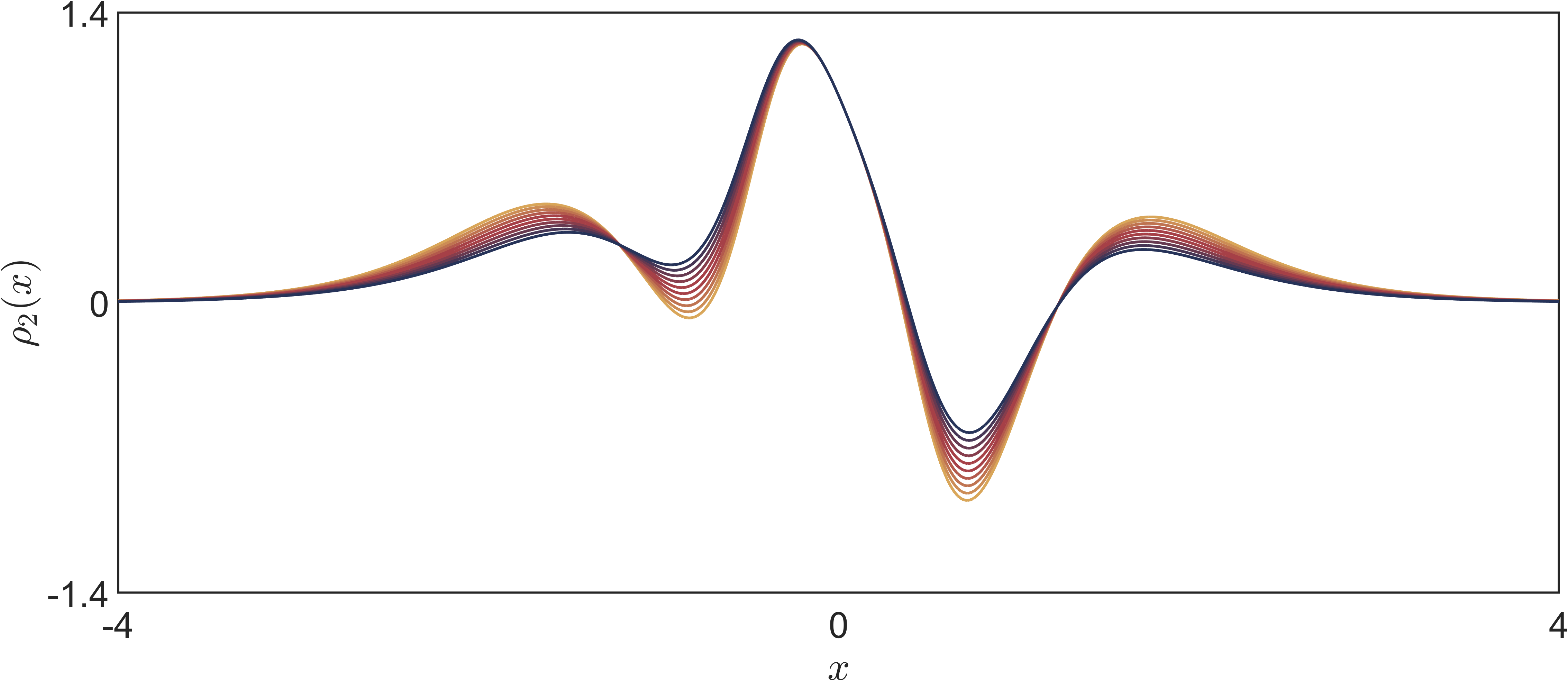}
    \caption{Energy density $\rho_{i}(x)$ given by Eqs. \eqref{rho4} and \eqref{rho5}. The parameter $\alpha$ is varied from $-5$ to $-3$, with the colors changing from lighter to darker.}
    \label{fig10}
\end{figure}

These results are plotted in Fig. \ref{fig10} for different values of $\alpha$. Despite the differences among the energy density profiles of the models presented so far, all the corresponding half-BPS solutions have the same total energy. This follows from the energy lower bound  \eqref{EB}, which only depends on the asymptotic behavior of the superpotential. For the $\phi^4$ models considered here, this yields $E_{B} = 4/3$.

\subsection{Two-Field Models}
The two-field kink-impurity framework is more involved than the single field case, not just because of the extra matter field and impurity, but also due to the extra coupling function $\mathcal{F}_{2}(\phi,\chi, \sigma_{1}, \sigma_{2})$. For this reason, we restrict our analysis to specific choices of these functions that lead to analytical solutions. For two matter fields, we use the notation where $\phi_{1} = \phi$ and $\phi_{2} = \chi$. Then, the general half-BPS equations in Eq. \eqref{eqBPS} reduce to
\begin{subequations}
\label{2fbps}
\begin{align}
    \phi^{\prime} &= \eta\PC{W_{\phi} + \sigma_{1}\mathcal{F}_{1\phi} + \sigma_{2}\mathcal{F}_{2\phi}},\\
    \chi^{\prime} &= \eta\PC{W_{\chi} + \sigma_{1}\mathcal{F}_{1\chi} + \sigma_{2}\mathcal{F}_{2\chi}}.
\end{align}
\end{subequations}
They form a system of two coupled first-order differential equations that, in general, cannot be solved analytically for arbitrary choices of $W$ and $\mathcal{F}_{i}$. In this sense, a convenient possibility is to consider
\begin{subequations}
\begin{align}
    \mathcal{F}_{1}\PC{\phi,\chi,\sigma_{1},\sigma_{2}} &= W\PC{\phi,\chi}\Gamma_{1}\PC{\sigma_{1}(x), \sigma_{2}(x)},\\
    \mathcal{F}_{2}\PC{\phi,\chi,\sigma_{1},\sigma_{2}} &= W\PC{\phi,\chi}\Gamma_{2}\PC{\sigma_{1}(x), \sigma_{2}(x)},
\end{align}
\end{subequations}
where $\Gamma_{1}$ and $\Gamma_{2}$ are arbitrary functions of the impurities. With this choice, the half-BPS equations simplify to
\begin{subequations}
\label{fosimp}
\begin{align}
    \phi^{\prime} &= \eta W_{\phi}\PC{1+\sigma_{1}\Gamma_{1} + \sigma_{2}\Gamma_{2}},\\
    \chi^{\prime} &= \eta W_{\chi}\PC{1+\sigma_{1}\Gamma_{1} + \sigma_{2}\Gamma_{2}}.
\end{align}
\end{subequations}
As previously implemented in the single field case, it is also possible to define a new function $\zeta(x)$ such that
\begin{equation}
\label{zeta}
    \zeta(x) = \int \PC{1+\sigma_{1}\Gamma_{1} + \sigma_{2}\Gamma_{2}}dx.
\end{equation}

The implications of introducing $\zeta(x)$, as well as its interpretation, are analogous to those discussed in the single field scenario. In particular, even when $\zeta(x)$ cannot be interpreted as a global coordinate, its composition with the standard solutions remains well defined. Therefore, if $\phi_{0}$ and $\chi_{0}$ solve the standard two-field BPS system, then $\phi(x) = \phi_{0}(\zeta(x))$ and $\chi(x) = \chi_{0}(\zeta(x))$ solve the half-BPS first-order system \eqref{fosimp}. Whenever $\zeta^{\prime}(x)\neq 0$, the first-order equations can be rewritten in the standard form
\begin{subequations}
\begin{align}
    \frac{d\phi_{0}}{d\zeta} &= \eta W_{\phi},\\
    \frac{d\chi_{0}}{d\zeta} &= \eta W_{\chi}.
\end{align}
\end{subequations}
If $\zeta^{\prime}$ vanishes only at isolated points, these relations remain valid there by continuous extension, although $\zeta$ ceases to define a regular local coordinate at those points.

\subsection*{B.1. Example}

\begin{figure}[!ht]
    \centering
\includegraphics[width=0.9\columnwidth,height=5cm]{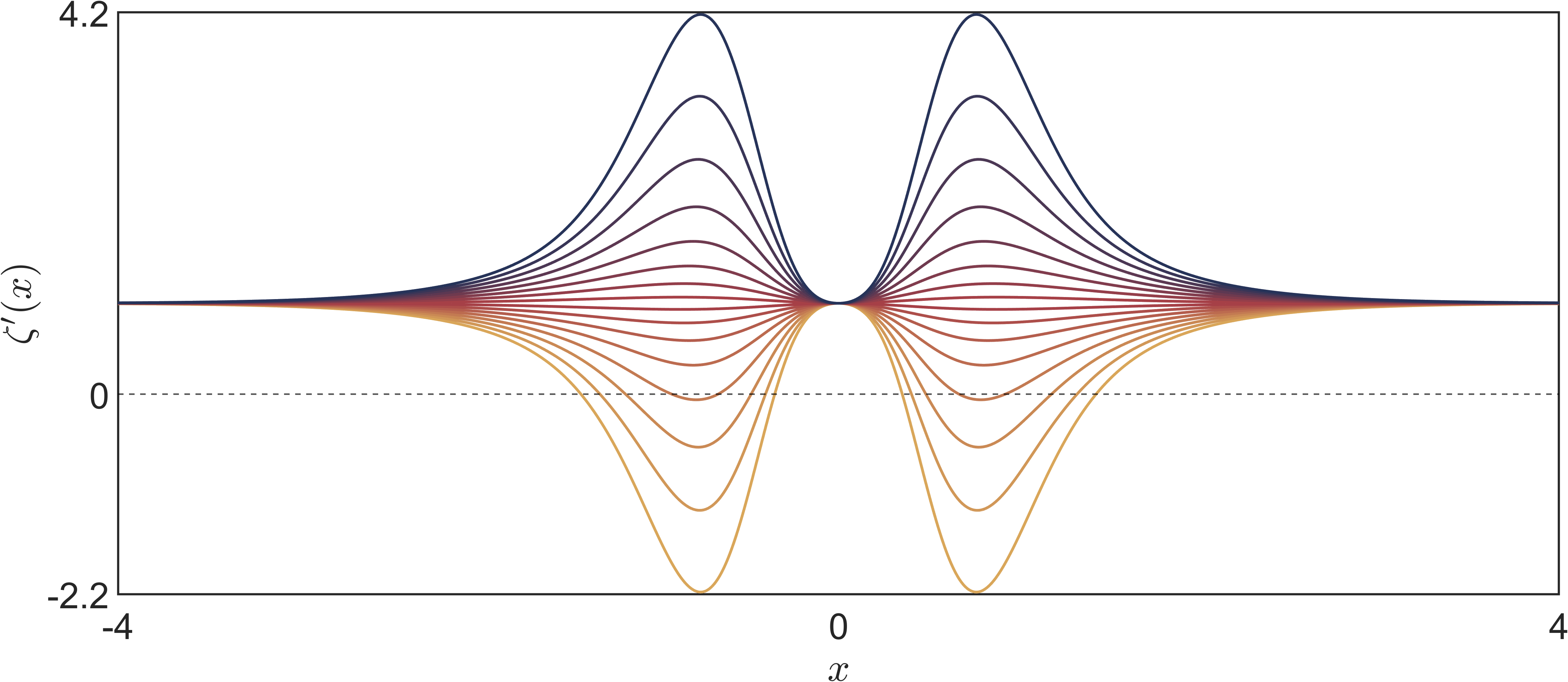}
\includegraphics[width=0.9\columnwidth,height=5cm]{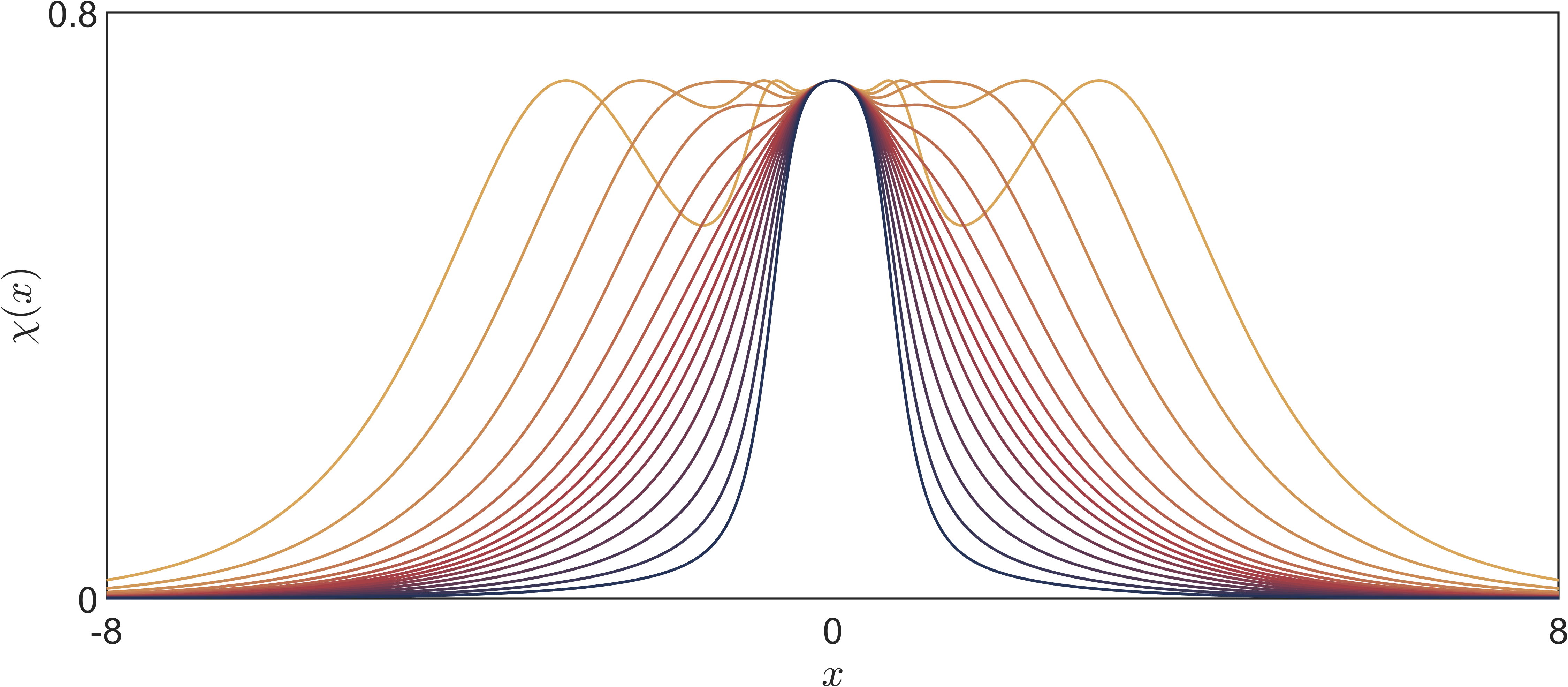}
\includegraphics[width=0.9\columnwidth,height=5cm]{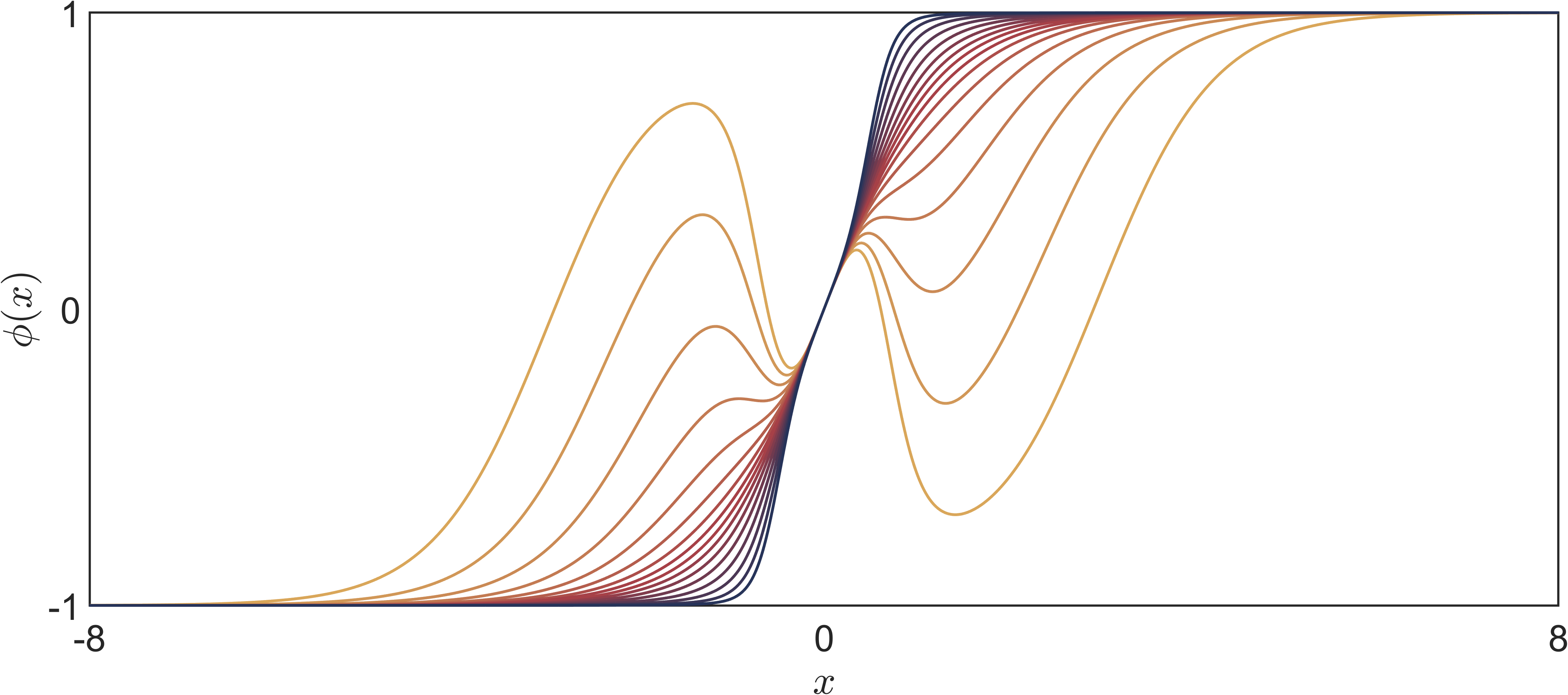}
    \caption{Profiles of $\zeta^{\prime}$ (top panel), $\chi(x)$ (middle panel), and $\phi(x)$ (bottom panel) for the BNRT model with $r=0.4$ and $\zeta(x)$ defined in Eq. \eqref{zeta1}. The parameter $\alpha\neq0$ increases from $-4$ to $4$ with the colors changing from lighter to darker.}
    \label{fig11}
\end{figure}

To illustrate this framework, we use the BNRT model introduced in Refs. \cite{BNRT1, BNRT2}. It is defined by the superpotential
\begin{equation}
\label{WBNRT}
    W(\phi,\chi) = \phi - \frac{1}{3}\phi^3 - r\phi\chi^2,
\end{equation}
which leads to the potential
\begin{equation}
    V(\phi,\chi) = \frac{1}{2}\PC{1-\phi^2-r\chi^2}^2 + 2r^2\phi^2\chi^2,
\end{equation}
where $r$ is a real parameter that controls the coupling of the fields. Taking $r>0$, the superpotential has two horizontal critical points, located at $\PC{\pm1,0}$, and two vertical, located at $\PC{0, \pm 1/\sqrt{r}}$. From the impurity-free case, it is known that a solution that interpolates from $(-1, 0)$ to $(1, 0)$ is given by
\begin{subequations}
\begin{align}
    \phi_{0}(x) &= \tanh(2rx),\\
    \chi_{0}(x) &= \sqrt{\frac{1-2r}{r}}\sech(2rx),
\end{align}
\end{subequations}
with $0<r<1/2$. The particular value of $r = 1/2$ is excluded from the interval since it degenerates the elliptic orbits on the field space into a straight line, trivializing one of the field solutions. In the supersymmetric framework developed so far, it is possible to obtain a family of analytical impurity-BNRT solutions. To do so, we need to choose $\sigma_{i}$ and $\mathcal{F}_{i}$ in a way that leads to an analytic new function $\zeta(x)$, defined in \eqref{zeta}. A possibility is to consider $\sigma_{1} = \sigma_{2} = \sech^2(x)$, $\Gamma_{1} = \alpha\PC{1-\sigma_{1}}$ and $\Gamma_{2} = \alpha^3\PC{1-\sigma_{1}}^2\sigma_{2}^2$, where $\alpha\in\mathbb{R}$ controls the strength of the coupling, with $\alpha = 0$ decoupling the impurities from the scalar fields. These choices lead to the function
\begin{equation}
\label{zeta1}
\begin{aligned}
    \zeta(x) &= x + \frac{\alpha}{3}\tanh^3(x) \\
    &+ \alpha^3\PC{\frac{\tanh^5(x)}{5} - 2\frac{\tanh^7(x)}{7} + \frac{\tanh^9(x)}{9}}.
\end{aligned}
\end{equation}
The corresponding solutions, for $\eta=1$, are given by $\phi(x)=\phi_0(\zeta(x))$ and $\chi(x)=\chi_0(\zeta(x))$, respectively. Their profiles, together with the behavior of $\zeta^{\prime}$ are displayed in Fig. \ref{fig11}, for $r = 0.4$ and for several different values of $\alpha$.

As shown in Fig. \ref{fig11}, some negative values of $\alpha$ cause $\zeta^{\prime}$ to cross the dashed zero line and change sign. As a consequence, $\zeta(x)$ becomes non-monotonic, leading to a non-monotonic solution profile $\phi(x)$. For values of $\alpha$ for which $\zeta^{\prime}$ remains positive, $\zeta(x)$ is monotonic, and the field configurations retain their usual profiles with no internal structures, becoming progressively localized as $\alpha$ increases. Since the BNRT model has been studied in different contexts, including Bloch branes \cite{Gomes} in braneworld scenarios and Bloch walls \cite{PRL01,PRL02}, the internal structures induced by the impurities provide additional motivation for investigating their possible implications in these contexts.

\begin{figure}[!ht]
    \centering
\includegraphics[width=0.9\columnwidth,height=5cm]{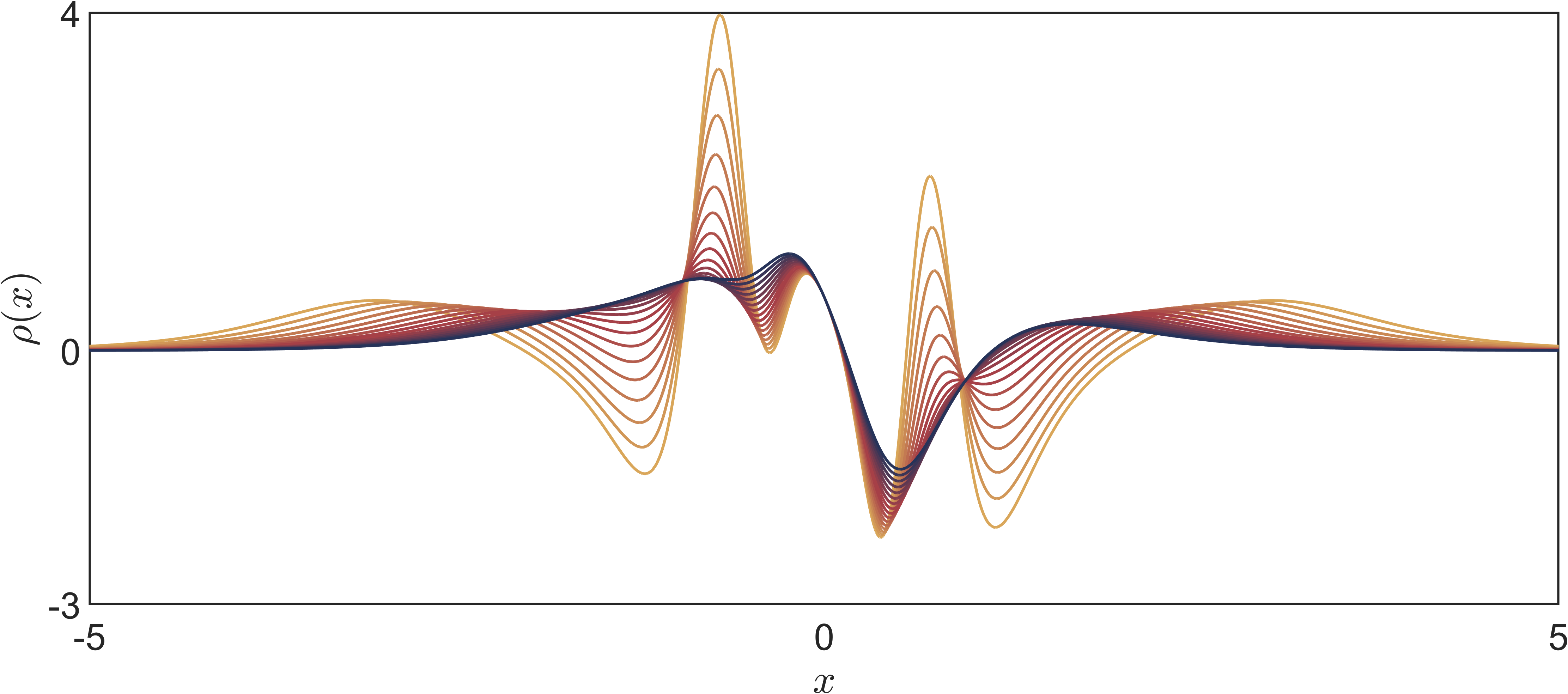}
\includegraphics[width=0.9\columnwidth,height=5cm]{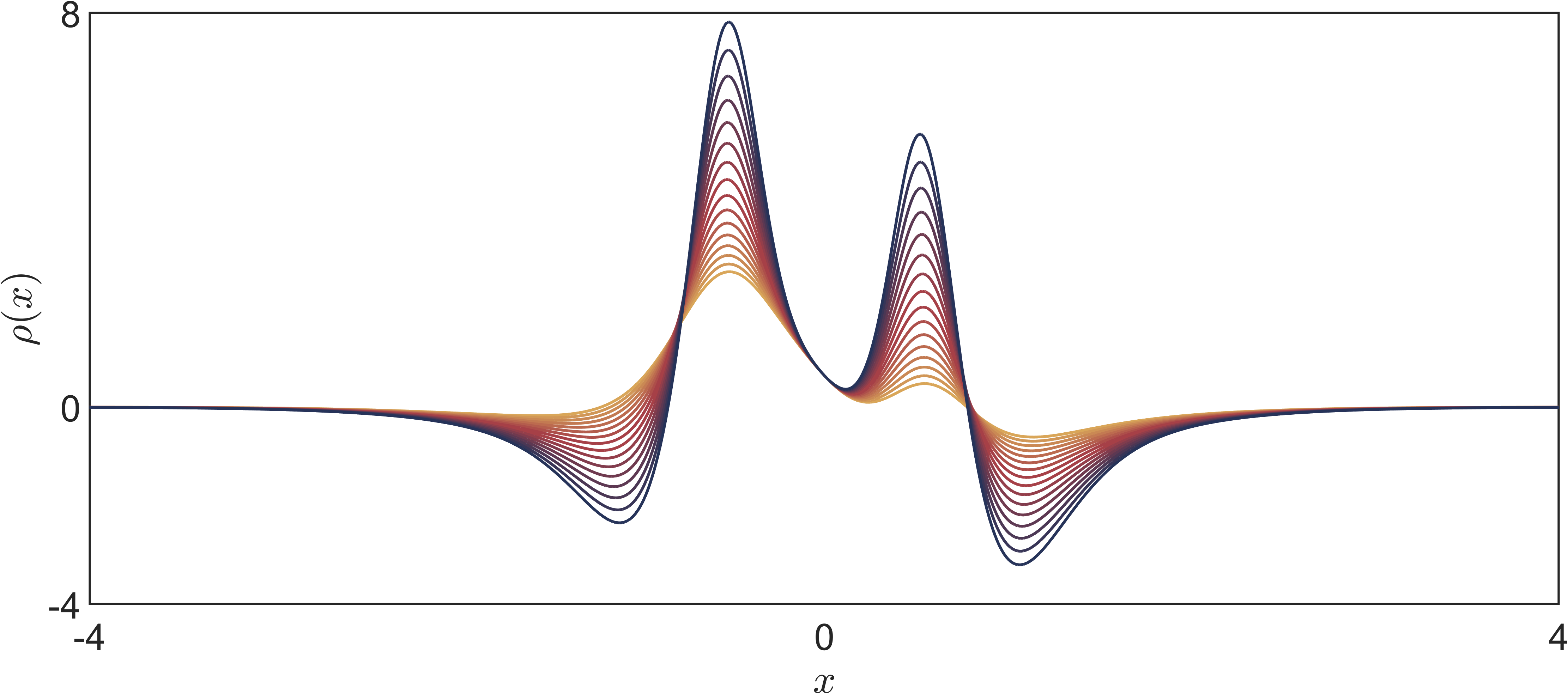}
    \caption{Energy density profiles given by Eq. \eqref{rhobnrt} for different values of $\alpha$. The parameter varies from $-4$ to $-2$ (top panel) and from $2$ to $4$ (bottom panel). In both panels, the curves change from lighter to darker colors as $\alpha$ increases.}
    \label{fig12}
\end{figure}

To express the energy density in a compact form, we substitute the half-BPS equations \eqref{eqBPS}, the function $\zeta(x)$ defined in Eq. \eqref{zeta}, and the specific choices $\mathcal{F}_{i} = W\Gamma_{i}$ into Eq. \eqref{rho}. This yields
\begin{equation}
    \rho(x) = \eta\frac{d}{dx}\PC{W\zeta^{\prime} + \frac{1}{2}\PC{\sigma_{1}^2 + \sigma_{2}^2}}.
\end{equation}
For the impurity-BNRT model, the energy density takes the explicit form
\begin{equation}
\label{rhobnrt}
\begin{aligned}
    \rho(&x) = \eta\frac{d}{dx}\bigg(\tanh(2r\zeta(x))\\
    &\;\;\times\!\PC{1 \!-\! \frac{1}{3}\tanh^2(2r\zeta(x)) \!-\! (1-2r)\sech^2\!(2r\zeta(x))}\\
    &\;\;\times\!\PC{\!1\!+\!\alpha\sech^2\!(x)\tanh^2\!(x) \!+\! \alpha^3\!\tanh^4\!(x)\sech^6\!(x)}\\
    &\;\;+ \sech^4(x)\bigg).
\end{aligned}
\end{equation}
Because the amplitude of the energy density is highly dependent on the parameter $\alpha$, Fig. \ref{fig12} displays its profiles for different intervals of this parameter, illustrating the distinct behavior for negative and positive values of $\alpha$.

It is worth emphasizing that, even within the range of $\alpha$ for which the function $\zeta(x)$ remains monotonic, the corresponding energy density still develops a nontrivial internal profile. This shows that the formation of internal structures in the energy density is not determined solely by the spatial behavior of the scalar fields, but also depends on their direct coupling to the impurities.

\subsection{Three-Field Models}

Let us now extend the framework by introducing a third real scalar field, together with an extra impurity and coupling function. Since this represents a further generalization of the two-field scenario, we also restrict our analysis to choices that yield analytical solutions. The notation used throughout this section is $\phi_{1} = \phi$, $\phi_{2} = \chi$ and $\phi_{3} = \psi$. Consequently, the half-BPS equations in Eq. \eqref{eqBPS} become
\begin{subequations}
\begin{align}
    \phi^{\prime} &= \eta\PC{W_{\phi} + \sigma_{i}\mathcal{F}_{i\phi}},\\
    \chi^{\prime} &= \eta\PC{W_{\chi} + \sigma_{i}\mathcal{F}_{i\chi}},\\
    \psi^{\prime} &= \eta\PC{W_{\psi} + \sigma_{i}\mathcal{F}_{i\psi}},
\end{align}
\end{subequations}
where repeated indices represent a summation over $i=1, 2$ and $3$. In general, this system of coupled first-order differential equations cannot be solved analytically. A particularly convenient choice is to consider
\begin{equation}
\label{3fchoice}
    \mathcal{F}_{i} = W(\phi,\chi,\psi)\Gamma_{i}(\sigma_{1}, \sigma_2, \sigma_3),
\end{equation}
where $\Gamma_{i}$ are arbitrary functions of the impurities. This choice changes the half-BPS equations to the form
\begin{subequations}
\begin{align}
    \phi^{\prime} &= \eta W_{\phi}\PC{1 + \sigma_{i}\Gamma_{i}},\\
    \chi^{\prime} &= \eta W_{\chi}\PC{1 + \sigma_{i}\Gamma_{i}},\\
    \psi^{\prime} &= \eta W_{\psi}\PC{1 + \sigma_{i}\Gamma_{i}}.
\end{align}
\end{subequations}
We note that the superpotential couples all three scalar fields. This is not a general requirement of the full framework, since the couplings could arise solely from the functions $\mathcal{F}_{i}$, but a consequence of the factorized choice \eqref{3fchoice}. As implemented in the one- and two-field scenarios, it is possible to define a new function $\tau(x)$
\begin{equation}
\label{tau}
    \tau(x) = \int \PC{1 + \sigma_{i}\Gamma_{i}}dx,
\end{equation}
which reparametrizes the half-BPS equations into their standard form. This prescription could be generalized to an arbitrary number of scalar fields. Also, the interpretation of this new function as a global or local coordinate follows a similar prescription of those already discussed in previous sections, for any number of scalar fields. However, obtaining explicit analytical solutions through this procedure requires the corresponding impurity-free multifield model to be analytically solvable, which is a challenging aspect for models with a higher number of fields.

\subsection*{C.1. Example}
An interesting three-field model is defined by the superpotential \cite{wot}
\begin{equation}
    W(\phi,\chi,\psi) = \phi -\frac{1}{3}\phi^3 - r\phi\PC{\chi^2+\psi^2} + rs\psi^2,
\end{equation}
which gives the potential
\begin{equation}
    V \!=\! \frac{1}{2}\!\PC{\!1\!-\!\phi^2\!-\!r\PC{\chi^2+\psi^2}\!}^2 \!+2r^2\phi^2\chi^2 \!+\! 2r^2\psi^2\PC{\!\phi-s\!}^2,
\end{equation}
where $r$ and $s$ are two real parameters, with $r$ encoding the strength of the coupling between all fields, since $r = 0$ decouples the three fields, resulting in a single $\phi^4$ model. For $r>0$ and $-1<s<1$, this model has 6 degenerate minima located at $(\pm1,0,0)$, $(0,\pm 1/\sqrt{r},0)$ and $(s, 0, \pm\sqrt{(1-s^2)/r})$, and supports several distinct topological solutions. In particular, imposing $\psi(x) = 0$ reduces the model to the BNRT case, defined by Eq. \eqref{WBNRT}. The above model is of interest for the entrapment of bosonic states in multifield models \cite{Brito,Casana}, to contribute to build analytical three-field cosmological models \cite{Joao,3field}, and for the construction of braneworld scenarios with a focus on the internal structure of the brane \cite{Lobao} and also, in Rastall gravity \cite{Liu}, to quote some possibilities of study.

To illustrate the prescription developed so far, we consider a specific solution whose orbit corresponds to a straight line connecting the minima $(0,1/\sqrt{r},0)$ and $(s, 0, \sqrt{(1-s^2)/r})$. The corresponding solutions are given by
\begin{subequations}
\begin{align}
    \phi_{0}(x) &= \frac{s}{2}\PC{1+\tanh(x/s)},\\
    \chi_{0}(x) & =  \frac{1}{2}\sqrt{\frac{1}{r}}\PC{1 - \tanh(x/s)},\\
    \psi_{0}(x) &=  \frac{1}{2}\sqrt{\frac{1}{r}(1-s^2)}\PC{1+\tanh(x/s)},
\end{align}
\end{subequations}
provided that $rs^2 = 1$. The subscript $0$ stands for the solution of the impurity-free theory. To obtain a family of analytical solutions in the supersymmetric impurity prescription, we need to define the new function $\tau(x)$ \eqref{tau} analytically. A simple choice is to eliminate the contributions associated with the second and third impurities by taking $\Gamma_{2} = \Gamma_{3} = 0$ or, equivalently, $\sigma_{2} = \sigma_{3} = 0$. In this case, $\tau(x)$ reduces to the function $\xi(x)$ defined in \eqref{xi}. Another possibility is to eliminate only the third impurity contribution, by taking $\Gamma_{3} = 0$ or $\sigma_{3} = 0$, in which  case $\tau(x)$ reduces to the function $\zeta(x)$ defined in Eq. \eqref{zeta}. For either choice, we could simply define the impurities and the coupling functions as in the corresponding one- or
two-field scenarios already discussed, thereby reproducing the same analytical functions and generating analytical three-field solutions. 

These reductions are not mandatory, since $\tau(x)$ may also be determined analytically while retaining nontrivial contributions from all three impurities. A possibility consists on the choices $\sigma_{1} = -\sech(x)\tanh(x)$, $\sigma_{2} = -\sech^2(x)\tanh(x)$, $\sigma_{3} = \sech^2(x)\tanh^2(x)$ and $\Gamma_{1} = \Gamma_{2} = \Gamma_{3} = \alpha$, which leads to the function
\begin{equation}
    \tau(x) = x + \alpha\PC{\sech(x) +\frac{\sech^2(x)}{2} + \frac{\tanh^3(x)}{3}}.
\end{equation}
As a consequence, the impurity three-field solution is given by $\phi(x) = \phi_{0}(\tau(x))$, $\chi(x) = \chi_{0}(\tau(x))$ and $\psi(x) = \psi_{0}(\tau(x))$. The main qualitative features of these solutions and their associated energy densities are analogous to those already examined, such as the formation of non-monotonic solutions and regions with negative energy densities, and for this reason we do not present additional plots for this case.

In any case, within this choice of $\mathcal{F}_{i}$ the energy density \eqref{rho} can be expressed in the simple form
\begin{equation}
    \rho(x) = \eta\frac{d}{dx}\PC{W\tau^{\prime}(x) + \frac{1}{2}\sigma_{i}^{2}}.
\end{equation}
Since we are dealing with localized impurities, the associated energy only depends on the asymptotic behavior of the superpotential, as defined in Eq. \eqref{EB}. This is also consistent with the definition of the new function $\tau(x)$, which has $\tau^{\prime}(x) = 1$ as $x\rightarrow\pm\infty$. As a consequence, the energy lower bound gives $E_{B} = \abs{s}(3-s^2)/3$.

\section{Outlook}
\label{sec4}

In this work, we developed a rigid supersymmetric spurionic formulation for localized structures in multifield scalar models in $1+1$ spacetime dimensions in the presence of impurities. The supersymmetric construction provides a controlled description of the bosonic half-BPS sector, in which the preserved projector selects the impurity-compatible first-order equations. Our approach was motivated by Ref.~\cite{Adam2}, where localized structures were studied in a supersymmetric environment, and by the recent work \cite{Lehum}, where spurion superfields were used to implement impurity backgrounds in a manifestly supersymmetric way. In the static bosonic sector, the formulation describes scalar fields coupled to localized impurity profiles, with the preserved supersymmetry controlling the BPS structure. This allowed us to derive, in a systematic way, the coupled first-order equations, energy density, boundary conditions, and Bogomol'nyi bound for impurity-deformed scalar configurations.

We then explored several representative models with one, two, and three real scalar fields, obtaining explicit solutions along with the corresponding  energy density profiles. In particular, we investigated models for which the first-order equations can be mapped onto those of the corresponding impurity-free theories through composition with a new function that, in some cases, could be interpreted as a reparametrization of the spatial coordinate. This mechanism is closely related to the geometric modification introduced before in Ref. \cite{Liao4}.

The presence of analytical solutions within this framework motivates several possible extensions. In particular, it would be of interest to investigate more general systems involving vector fields in two or three spatial dimensions \cite{sup1,sup2,sup3,sup4,BLM}. Such extensions may reveal new possibilities in which supersymmetric impurity models can be further considered. The inclusion of the spurionic contribution seems to be an interesting possibility, and we are now investigating other modifications, in particular, the case with two fields, with the second field adding a new interaction that modifies the kinetic contribution of the first field in a way similar to the study previously developed in Refs. \cite{Liao4,Andrade,Simas}. We hope to report on them in the near future.

\acknowledgments{This work is supported by the Brazilian agencies Conselho Nacional de Desenvolvimento Cient\'ifico e Tecnol\'ogico (CNPq), grants Nos. 402830/2023-7 (DB), 303469/2019-6 (DB), 404310/2023-0 (ACL) and \ 301256/2025-0 (ACL), and Coordenação de Aperfeiçoamento de Pessoal de Nível Superior (CAPES), Grant No. 88887.132514/2025-00 (GSS)}.

{\textbf {Data availability statement:}} 
All data that support the findings of this study are included within the article.

{\textbf{Authors statement:}} Authors have no relevant financial or non-financial interests to disclose.


\end{document}